\documentclass[journal,11pt,draftclsnofoot,onecolumn]{IEEEtran}

\usepackage{cite}
\usepackage{amssymb}
\usepackage[pdftex]{graphicx}
\graphicspath{{./}{figures/}}
\usepackage[table]{xcolor}
\usepackage{amsmath}
\usepackage{amssymb}
\usepackage{url}
\usepackage{setspace}
\usepackage{fixltx2e}
\usepackage{floatrow}
\newfloatcommand{capbtabbox}{table}[][\FBwidth]
\usepackage{blindtext}
\def\eg{\emph{e.g. }}
\def\ie{\emph{i.e. }}
\def\etal{\emph{et al. }}

\def\astoday{as the date of 04/10/2012}

\usepackage{algorithm}
\usepackage{algorithmic}
\usepackage{rotating}
\usepackage{multirow}
\definecolor{color1}{RGB}{252,213,181}
\definecolor{color2}{RGB}{119,147,60}
\definecolor{color3}{RGB}{149,179,215}
\definecolor{color4}{RGB}{99,37,35}
\begin{document}
\title{Sudoku Associated Two Dimensional Bijections for Image Scrambling}
\author{Yue~Wu,~\IEEEmembership{Student Member,~IEEE,}
        {Sos~Agaian},~\IEEEmembership{Senior Member,~IEEE,}\\
        and Joseph~P.~Noonan,~\IEEEmembership{Life Member,~IEEE,}
\thanks{Yue Wu and Joseph P. Noonan is with the Department of Electrical and Computer Engineering, Tufts University, Medford, MA 02155, United States; e-mail: ywu03@ece.tufts.edu.}
\thanks{Sos Agaian is with the Electrical and Computer Engineering, University of Texas at San Antonio, San Antonio, TX 78249, United States.}
}

\markboth{A Paper Draft Submitted to IEEE Transactions on Multimedia}%
{Wu \etal}
\maketitle

\begin{abstract}
Sudoku puzzles are now popular among people in many countries across the world with simple constraints that no repeated digits in each row, each column, or each block. In this paper, we demonstrate that the Sudoku configuration provides us a new alternative way of matrix element representation by using block-grid pair besides the conventional row-column pair. Moreover, we discover six more matrix element representations by using row-digit pair, digit-row pair, column-digit pair, digit-column pair, block-digit pair, and digit-block pair associated with a Sudoku matrix. These parametric Sudoku associated matrix element representations not only allow us to denote matrix elements in secret ways, but also provide us new parametric two-dimensional bijective mappings. We study these two-dimensional bijections in the problem of image scrambling and propose a simple but effective \textit{Sudoku Associated Image Scrambler} only using Sudoku associated two dimensional bijections for image scrambling without bandwidth expansion. Our simulation results over a wide collection of image types and contents demonstrate the effectiveness and robustness of the proposed method. Scrambler performance analysis with comparisons to peer algorithms under various investigation methods, including human visual inspections, gray degree of scrambling, autocorrelation coefficient of adjacent pixels, and key space and key sensitivities, suggest that the proposed method outperforms or at least reaches state-of-the-art. Similar scrambling ideas are also applicable to other digital data forms such as audio and video.

\end{abstract}

\begin{IEEEkeywords}
Sudoku Matrix, Matrix Element Representation, Two Dimensional Bijection, Image Scrambling
\end{IEEEkeywords}
\IEEEpeerreviewmaketitle
\section{Introduction}
In the new century, business news about conventional industries and companies keep reminding us how much impact the digital revolution bring to us and to the world. \textit{United States Postal Service} is struggling hard to survive these days and have already closed hundreds of facilities and sites, because paper letters are not used as often as they were in the last century after email and other electronic communications dominates the way that people communicate with each other. \textit{Borders}, an international book and music retailer once pioneered in the kind of large supermarket style bookstores, failed to survive under the impacts from online and e-books and has already bankrupted in the early of the year. Similar stories also happened to \textit{Eastman Kodak company}, commonly known as \textit{Kodak}, an American multinational imaging and photographic equipment, materials and service company that once dominated the photographic film industry for many years and invent and built the first digital camera as early as in 1975 \cite{prel2009visual}, but now was bankrupted due to its slowness in transitioning to digital photography. As emails, digital images, digital books and other digital data carriers play an important role in people's life, the demand on the secure storage and transmission on digital data become a problem must be solved.


Digital images, one typical type of two dimensional data, are considered to contain a huge amount of information, for example, a family photo might tell not only who are in this family and how they look like, but rough ages for each member and maybe their healthy conditions; a diagnosis CT image might tell a doctor whether the patient is healthy or not and if he/she is sick how bad he/she is; and a satellite photo might give information whether the interested region is under constructions and what these constructions could be. Because the information contained in a digital image and the information might be inferred beyond a digital image, it is very important to protect these information from any unauthorized use. One way of protecting digital images is called image scrambling, which disorders pixel relationship in the original image so that the scrambled image with rearranged pixels become unintelligent and unrecognizable.

Mathematically speaking, all image scrambling algorithms rely on a two dimensional bijection, which maps pixels in the original image to rearranged pixels in the scrambled image. According to the generation of a two dimensional bijection used in a scrambler, we can classify a scrambler into groups: chaos based \cite{Ye2010,Zhang2011}, spatial transform based \cite{ye2007image,lin2011hybrid}, matrix decomposition based \cite{VanScrambler} and cellular automata based \cite{ye2008novel,Abu}. According to the way that an image scrambler uses a two dimensional bijection, we can roughly classify it into two groups: conventional scrambling \cite{VanScrambler} and total scrambling \cite{Abu,ye2007image,ye2008novel,Ye2010}. Conventional scrambling consider each pixel an unit and only shuffles pixel positions, while total scrambling consider each bit of a pixel an unit and shuffles pixel bit positions and pixel positions. Generally speaking, the total scrambling scheme is considered more secure than the conventional scheme because image histograms have been changed after total scrambling, but stays unchanged after conventional scrambling. Moreover, image scrambling are an essential component for advanced data protection techniques, for example, image encryption \cite{3DCat,3DBaker,wuSPIE,zhu2011chaos,Ruisong2011,Zhou2012PFibonacci,Patidar2011,Liu2011,Fu2011,wuLogistic} and watermarking \cite{5331518,5656731,5555253,6057940}.

In this paper, we extend our work in \cite{wuSPIE,wuIEEE} and introduce new ways of constructing two dimensional bijections for digital image scrambling using Sudoku matrices. We demonstrate that a $N\times N$ Sudoku matrix can be used to define six new parametric matrix element representations for a $N\times N$ matrix. Consequently, we are able to construct Sudoku associated two dimensional bijection by mapping one matrix element representation to the other. Moreover, we demonstrate that Sudoku associated two dimensional bijections actually performs scrambling in a guarantee way. For example, the Sudoku associated two dimensional bijection mapping from the row-digit pair to the row-column pair is to scramble image pixels in such a way that none two pixels originally lies in the same column will be in different column after scrambling, and the Sudoku associated two dimensional bijection mapping from the digit-column pair to the row-column pair is to scramble image pixels in such a way that none two pixels originally lies in the same row will be in different row after scrambling. In addition, we propose a total scrambling scheme \textit{Sudoku Associated Image Scrambler} only using on these Sudoku associated two dimensional bijections for scrambling. Simulation results on various image types and datasets demonstrate the effectiveness and robustness of the proposed method. Visual and Analytical comparisons to peer algorithms suggest that the \textit{Sudoku Associated Image Scrambler} outperforms or reach state-of-the-art of image scrambling. Statistical testing results also support that \textit{Sudoku Associated Image Scrambler} successfully decorrelates pixels in the original image to random-like.

The commonly used symbols and notations in this article are given in Table \ref{tab:sym}. In the rest of the article, Section II gives a brief background about Sudoku matrices; Section III discusses the Sudoku associated matrix element representations and two dimensional bijections; Section IV proposes the \textit{Sudoku Associated Image Scrambler} using Sudoku associated two dimensional bijections; Section V shows extensive simulation results and compares the performance of \textit{Sudoku Associated Image Scrambler} with peer algorithms in detail; Section VI concludes the article.

\begin{table}
\caption{Symbol and Notation Usage}\label{tab:sym}
\scriptsize
\centering
\begin{tabular}{r|l}
  \hline\hline
  \textbf{Symbol} & \textbf{Description}\\\hline
  $N$ & a squared integer\\
  $I(.)$ & an indication function on range $\{0,1\}$\\
  $r$ & row index of an element in a matrix\\
  $c$ & column index of an element in a matrix\\
  $b$ & block index of an element in a Sudoku matrix\\
  $g$ & grid index of an element in a Sudoku matrix\\
  $d$ & digit index of an element within a Sudoku matrix\\
  $X_{r,c}$ & a pixel/element located at the intersection of the $r$th row and $c$th column of image/matrix $X$ \\
  $S$ & a Sudoku matrix\\
  $\mathbb{I}$ & the set of natural numbers from $1$ to $N$\\
  $f$ & a two dimensional bijection on $\mathbb{I}\times \mathbb{I}$\\
  $f^{-1}$ & the inverse bijection of $f$\\
  $f_{R\rightarrow R'}$ & a two dimensional bijection mapping from a representation $R$ to $R'$ \\
  $f_{R\leftarrow R'}$ & a two dimensional bijection mapping from a representation $R'$ to $R$ \\
  $fix(.)$ & rounding function towards to zero\\
  $rem(.)$ & remainder function\\
  $F$ & a fixed matrix element representation pair\\
  $P$ & a matrix element parametric representation pair\\
  $R_F$ & a fixed matrix element representation using pair $F$\\
  $R_P^S$ & a matrix element representation using pair $P$ with reference Sudoku $S$\\
  $\circ$ & the function composition symbol\\
  $W$ & the width of an image\\
  $H$ & the height of an image\\
  $T$ & the number of pixels of an image\\
  $K$ & a scrambling key\\
  $\Gamma(.)$ & the Gamma function\\
  $|.|$ & the absolute value function \\
  $g(.)$ & the Student's t-distribution function\\
  \hline\hline
\end{tabular}
\end{table}

\section{Background}
The name \textit{Sudoku} is the abbreviation of the Japanese '\textit{Sunji wa dokushin ni kagiru}', which means 'single number'. Conventionally, Sudoku refers to a number-based puzzle, consisting of $9\times 9$ grids\footnotemark[1] divided into nine $3\times 3$ blocks \cite{SudokuSci}. The objective is to complete the grids using digits ranging from $1$ to $9$, in a manner that there are no repeated digits in any single row, column and block of the overall puzzle.

\footnotetext[1]{in some literature, this $3\times 3$ block is referred to as a \textit{box}, or a \textit{square}.}

\begin{figure}[ht!]
    \begin{minipage}[b]{0.45\linewidth}
      \centering
     \centerline{\includegraphics[width=.95\linewidth]{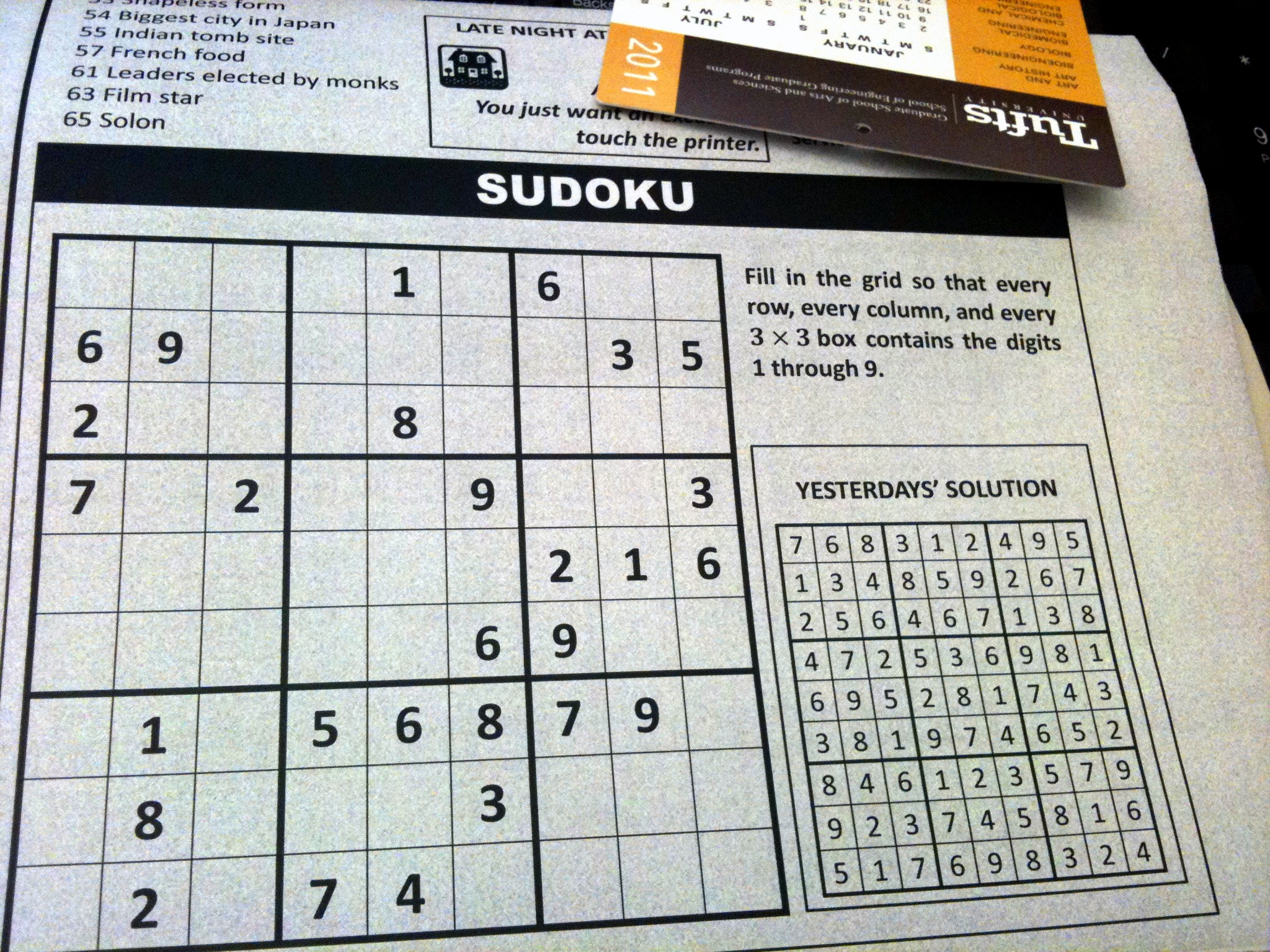}}
     \centerline{\scriptsize(a)}
      \vspace{0.12cm}
    \end{minipage}\hfill
    \hspace{-20pt}
    \begin{minipage}[b]{0.43\linewidth}
      \centering
     \centerline{\includegraphics[width=.75\linewidth]{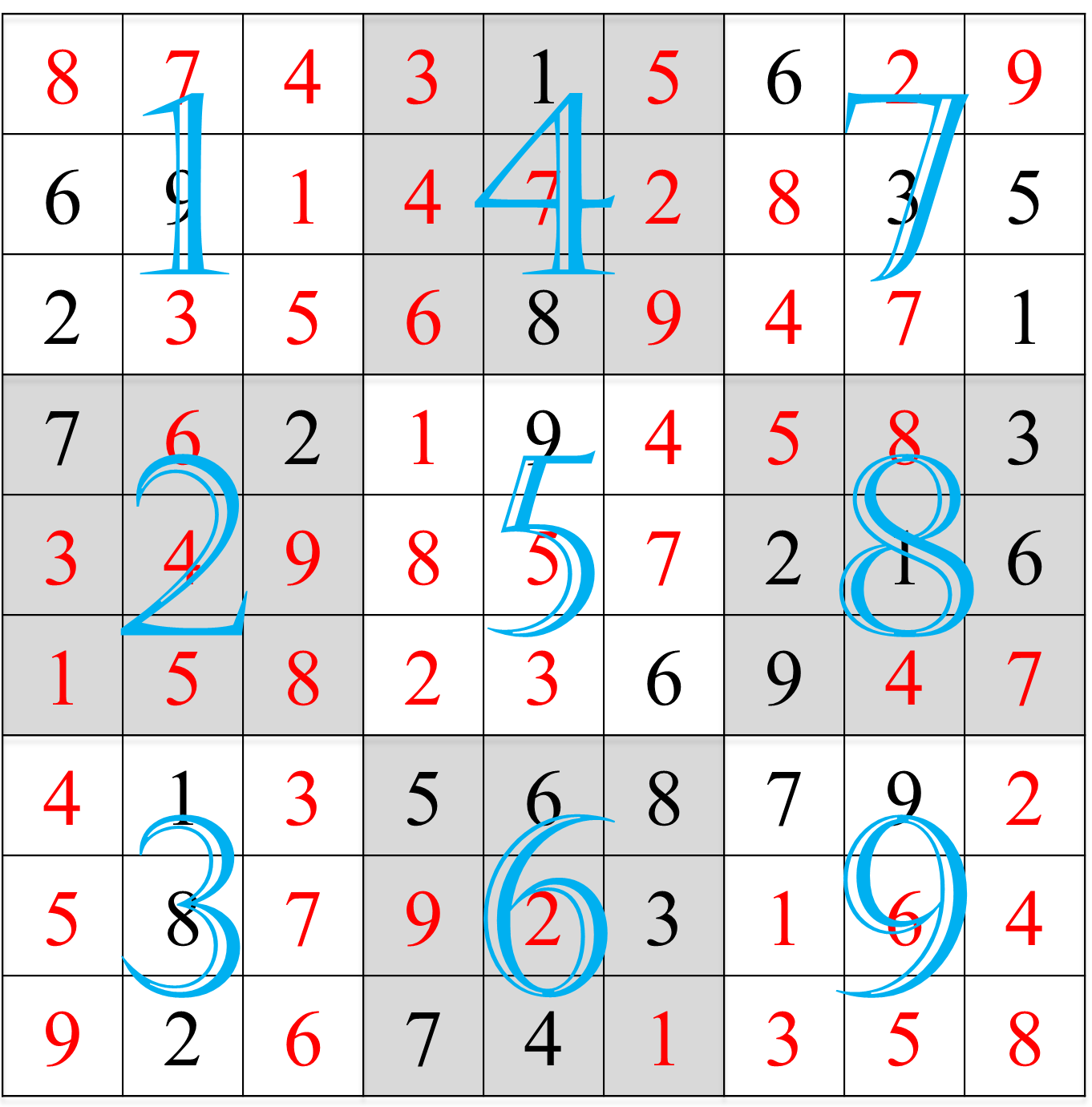}}
     \centerline{\scriptsize(b)}
      \vspace{0.12cm}
    \end{minipage}\hfill
    \caption{A newspaper style Sudoku puzzle and its solution, (a) a Sudoku puzzle, and (b) its solution}
    \label{fig:Sudoku_newspaper}
\end{figure}

Fig. \ref{fig:Sudoku_newspaper} shows a Sudoku puzzle in a newspaper and its solution. The Sudoku hints given \ref{fig:Sudoku_newspaper}-(a) are identified by black colored numerals; the nine Sudoku block indices are identified by the large blue colored numerals ranging from $1$ to $9$; and the blank Sudoku elements in \ref{fig:Sudoku_newspaper}-(a) are identified by red colored numerals in Fig. \ref{fig:Sudoku_newspaper}-(b), respectively. This is a conventional Sudoku puzzle, with a $9\times 9$ size, to be filled with digits ranging from $1$ to $9$, and divided in square blocks of size $3\times 3$.

\begin{figure}
\centering
\scriptsize
    \begin{minipage}[c]{0.45\linewidth}
      \centering
        \centerline{\includegraphics[width=.95\linewidth]{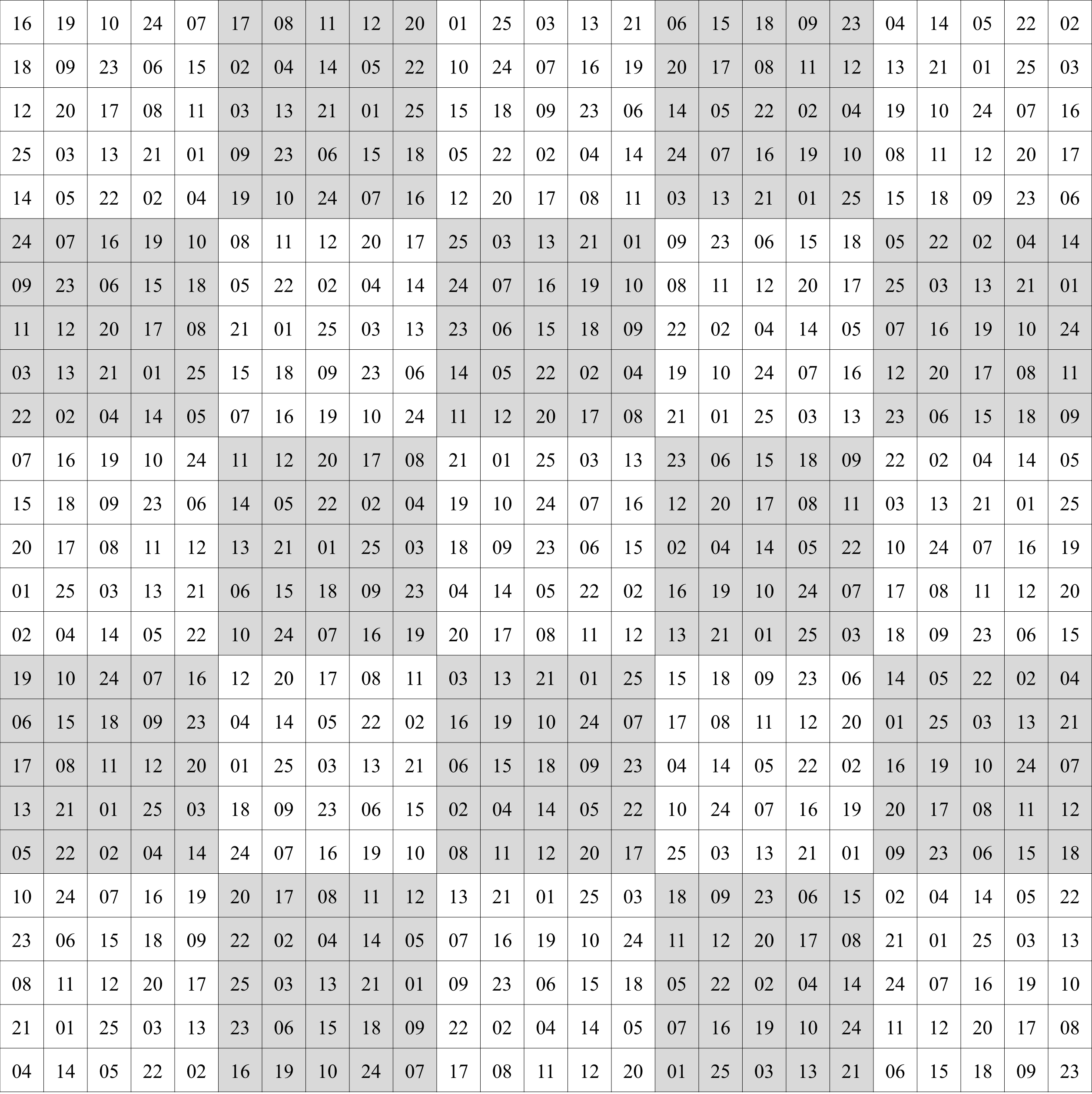}}
      \centerline{\scriptsize (a)}
    \end{minipage}\hfill
        \begin{minipage}[c]{0.54\linewidth}
      \centering
        \centerline{\includegraphics[width=.95\linewidth]{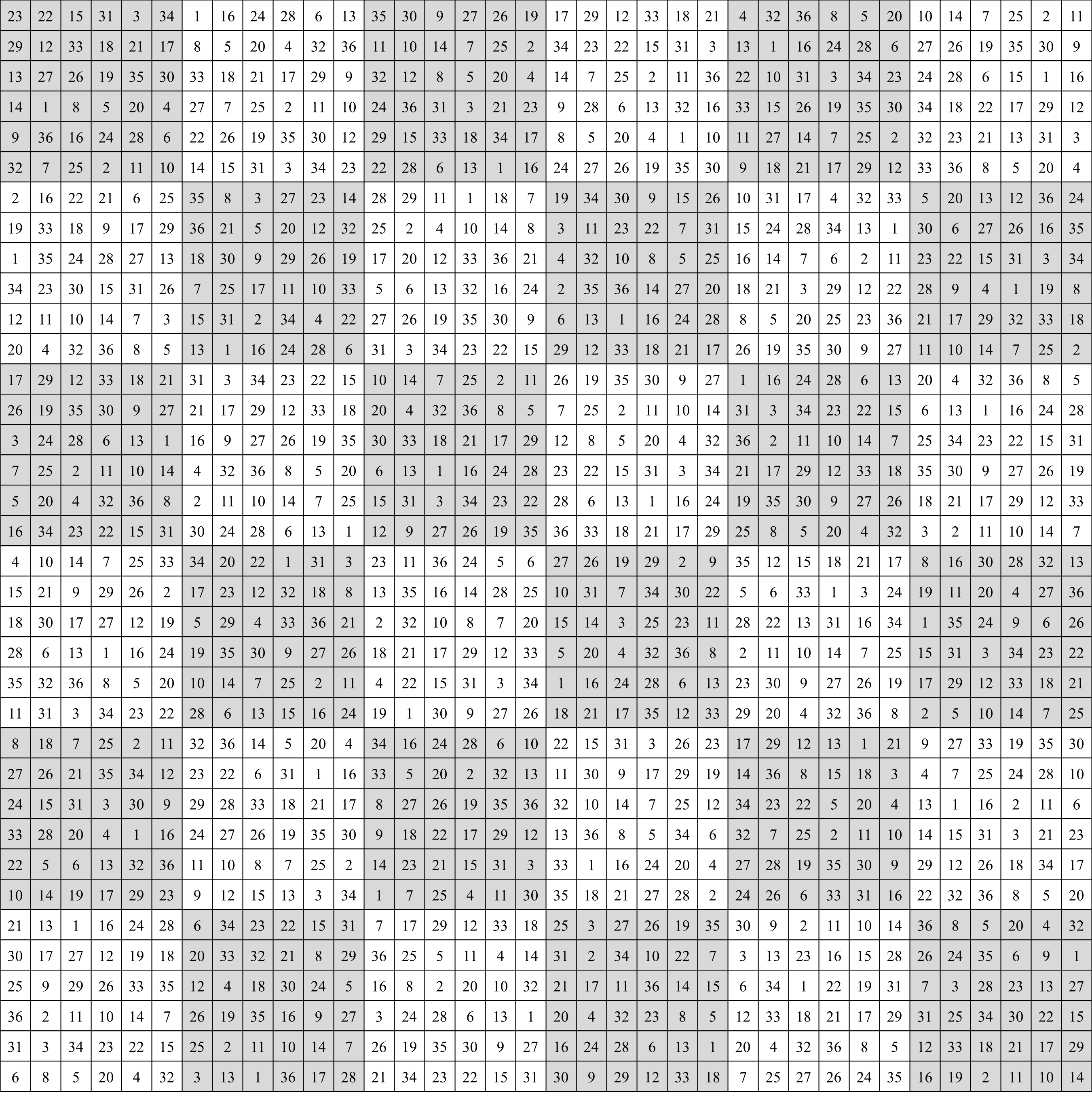}}
      \centerline{\scriptsize (b)}
    \end{minipage}\hfill
      \caption{Large size Sudoku matrices, (a) a $25\times 25$ Sudoku, and (b) a $36\times 36$ Sudoku}\label{Fig:Sudoku_matrices}
\end{figure}

Throughout of this paper, we are interested in a Sudoku solution instead of a Sudoku puzzle, because multiple Sudoku puzzles can be made with respect to a single Sudoku solution, and it is the Sudoku solution that satisfies all digit constraints along rows, columns and blocks. Specifically, we consider a Sudoku solution from the point of view of matrix and thus call it a Sudoku matrix. It is noticeable that one can easily extend the conception of a $9\times 9$ Sudoku matrix for other squared sizes, \eg $4\times 4$, $16\times 36$, $25\times 25$ etc. Fig. \ref{Fig:Sudoku_matrices} shows sample results for large size Sudoku matrices. Literally, we are therefore able to define a $N\times N$ Sudoku matrix with $N=n^2$ as shown in Def. 1.

\noindent \textbf{Definition 1 (Sudoku Matrix):} A $N\times N$ matrix is a Sudoku matrix if and only if its elements satisfy all three constraints that matrix elements in any row, and in any column, and in any $n\times n$ block contains exact $N$ digits from $1$ to $N$.

A $N\times N$ Sudoku matrix conforming Def. 1 has the properties including, but not limited to the three listed below
\begin{itemize}
  \item The $N$ elements within each row of this Sudoku matrix is a permutation of the nature number sequence $\{1,2,\cdots,N\}$
  \item The $N$ elements within each column of this Sudoku matrix is a permutation of the nature number sequence $\{1,2,\cdots,N\}$
  \item The $N$ elements within each block of this Sudoku matrix is a permutation of  the nature number sequence $\{1,2,\cdots,N\}$
  \item A subset of $N\times N$ Sudoku matrices can be perimetrically generated \cite{wuSPIE}
\end{itemize}
It is obvious that these three properties are directly from the definition of a Sudoku matrix.

\section{Methodology}
In this section, we mainly explore the possible matrix representations associated with a Sudoku matrix. And we show that element-wise mapping between any of two representations is actually a bijection, which ensures to shuffle and restore image pixels in an easy way.
\subsection{A Sudoku Matrix Definition}
Mathematically, we can define a $N\times N$ squared size Sudoku matrix via an indicator function $I$ as follows:
\begin{equation}\label{eq:Sudoku}
    I_X(r,c,d) = \left\{\begin{array}{l}1, \textrm{ if } X_{r,c} = d\\0, \textrm{ Otherwise}\end{array}\right.
\end{equation}
where $r,c$ and $d$ denote the row index, column index, and digit index of a matrix element in matrix $X$; and $X_{r,c}$ denotes the element at the intersection of the $r$th row and $c$th column in $X$. Let $\mathbb{I} = \{1,2,\cdots,N\}$ be the index set. Then

\noindent\textbf{Definition 2 (Sudoku Matrix):} A $N\times N$ matrix $S$ is a Sudoku matrix if and only if
\begin{itemize}
  \item For arbitrary $r,d\in \mathbb{I}$, we have $\sum\limits_{r=1}^{N}{I_S(r,c,d) = 1}$
  \item For arbitrary $c,d\in \mathbb{I}$, we have $\sum\limits_{c=1}^{N}{I_S(r,c,d) = 1}$
  \item For arbitrary $b,d\in \mathbb{I}$, we have $\sum\limits_{g=1}^{N}{I_S(b_r^g,b_c^g,d) = 1}$
\end{itemize}
where $r,c,b$ and $d$ denotes the row index, column index, block index and digit index, respectively; $b_r^g$ and $b_c^g$ denote the row index and the column index of the $g$th grid in the $b$th block.

In particular,  the row index $r$ and the column index $c$ of the $g$th grid in the $b$th block are defined as follows,
\begin{equation}\label{eq.blockRC}
    f_{(b,g)\rightarrow(r,c)}:\left\{\begin{array}{l}b_r^e = rem(b-1,n)\cdot n+rem(g-1,n)+1\\b_c^g=fix(b-1,n)\cdot n+fix(g-1,n)+1\end{array}\right.
\end{equation}
with the rounding function towards to zero $fix(.)$, namely $fix(p,q) = \lfloor p/q\rfloor$ and the remainder function $rem(.)$ , namely $rem(p,q) = p-fix(p,q)\cdot q$. In this way, we define the two-dimensional mapping from the Sudoku representation $R_{(b,g)}$ to the conventional representation $R_{(r,c)}$. It is noticeable that this is a bijective mapping, and we can write the inverse mapping as
\begin{equation}\label{eq.RCblock}
    f_{(r,c)\rightarrow(b,g)}:\left\{\begin{array}{l}g = rem(r-1,n)\cdot n+rem(c-1,n)+1\\b=fix(r-1,n)\cdot n+fix(c-1,n)+1\end{array}\right.
\end{equation}

Although mathematically the bijective mapping between representation $R_{(r,c)}$ and representation $R_{(b,g)}$ can be explicitly written as Eqs. \eqref{eq.blockRC} and \eqref{eq.RCblock}, such bijective mapping is actually nothing but to link two representations with respect to each matrix element.

\begin{figure}[h]
\centering
\scriptsize
\begin{minipage}[c]{0.65\linewidth}
    \begin{minipage}[c]{0.33\linewidth}
      \centering
        \centerline{\includegraphics[width=.95\linewidth]{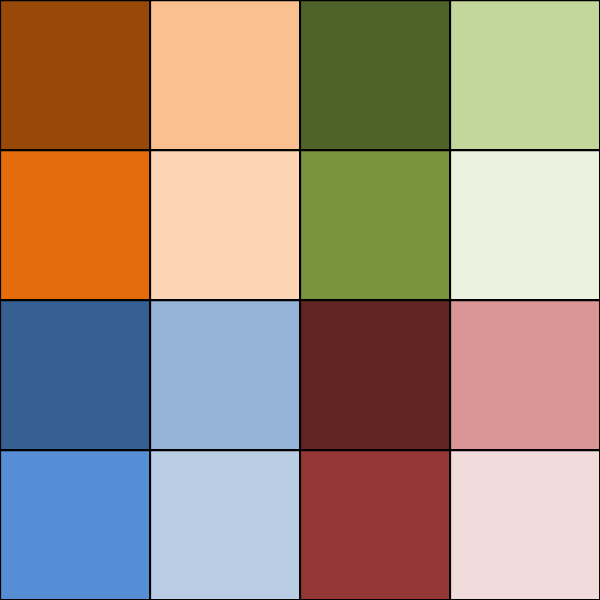}}
      \centerline{(a)}
    \end{minipage}\hfill
        \begin{minipage}[c]{0.33\linewidth}
      \centering
        \centerline{\includegraphics[width=.95\linewidth]{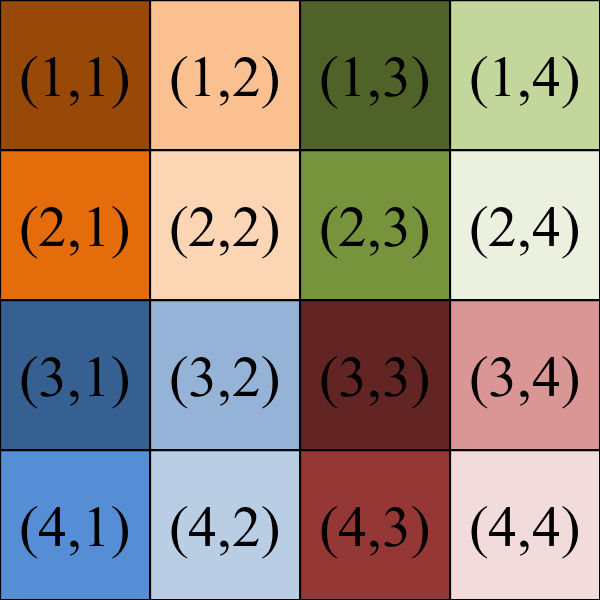}}
      \centerline{(b)}
    \end{minipage}\hfill
        \begin{minipage}[c]{0.33\linewidth}
      \centering
        \centerline{\includegraphics[width=.95\linewidth]{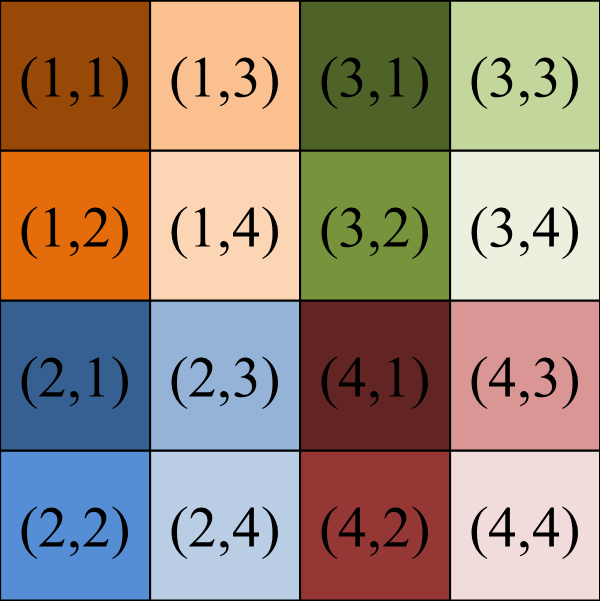}}
      \centerline{(c)}
    \end{minipage}\hfill
    \end{minipage}\hfill
     \begin{minipage}[c]{0.95\linewidth}
      \centering
        \centerline{\includegraphics[width=.9\linewidth]{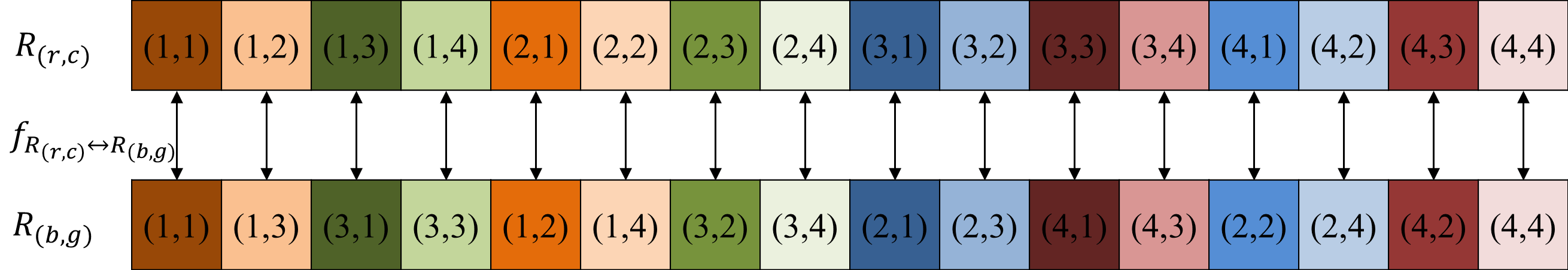}}
      \centerline{(d)}
    \end{minipage}\hfill
  \caption{Matrix element representation example; (a) a $4\times 4$ matrix with each element identified in a unique color; (b) the conventional row-column representation $R_{(r,c)}$; (c) the Sudoku representation $R_{(b,g)}$; and (d) the bijective mapping between matrix element representation $R_{(r,c)}$ and $R_{(b,g)}$. }\label{fig:rcbk}
\end{figure}

Fig. \ref{fig:rcbk} shows an example of matrix element representations $R_{(r,c)}$ and the $R_{(b,g)}$  for a $4\times 4$ matrix. As can be seen, the actual mapping from representation $R_{(r,c)}$ to representation $R_{(b,g)}$ defined in Eq. \eqref{eq.blockRC} can be simply described to map a $(r,c)$ pair in Fig. \ref{fig:rcbk}-(b) to its corresponding $(b,g)$ pair in Fig. \ref{fig:rcbk}-(c) with the same color. Similarly, the actual mapping from representation $R_{(b,g)}$ defined in Eq. \eqref{eq.RCblock} to representation $R_{(r,c)}$  is to map a $(b,g)$ pair in Fig. \ref{fig:rcbk}-(c) to its corresponding $(r,c)$ pair in Fig. \ref{fig:rcbk}-(b) with the same color. Consequently, the bijection between these two representations are shown in  Fig. \ref{fig:rcbk}-(d).

So far we have two matrix element representations, namely row-column pair representation $R_{(r,c)}$ and block-element pair representation $R_{(b,g)}$. To differentiate from those representation associated with one or more Sudoku matrices, we call these two representations \textit{fixed representations} because they are fixed rather than parametric and use symbol $F$ to denote a fixed representation pair, \ie $F\in \{(r,c), (b,g)\}$. 

\subsection{Sudoku Associated Matrix Element Representations}
Considering the task of digital image scrambling, the two dimensional bijection $f_{R_{(r,c)}\leftarrow R_{(b,g)}}$ and $f_{R_{(r,c)}\rightarrow R_{(b,g)}}$ are inappropriate in the sense that such bijection is fixed. With the help of a Sudoku matrix, however, we are able to easily define two dimensional parametric bijections between two matrix element representations.

Due to the digit constraints in each row, each column and each block within a $N\times N$ Sudoku matrix $S$, we are able to create parametric matrix element representations for a $N\times N$ matrix using the reference Sudoku matrix $S$. Specifically, we are able to have six parametric matrix element representations using row-digit pair $(r,d)$, digit-row pair $(d,r)$, column-digit pair $(c,d)$, digit-column pair $(d,c)$, block-digit pair $(b,d)$ and digit-block pair $(d,b)$. Intuitively, one can convince him/herself that matrix elements can be represented with the $(r,d)$ pair, because each matrix can be decomposed with respect to rows, and each element in a row is with a unique digit associated with a reference Sudoku in its row, implying that one can unique locate a matrix element whenever he/she knows the row index $r$, the digit index $d$ and the reference Sudoku matrix $S$. Similarly, it is not difficult to see that other listed Sudoku associated pairs are also valid matrix element representations. We denote these Sudoku associated matrix element representations as $R^S_P$ where $S$ is the associated Sudoku matrix and $P$ denotes a Sudoku associated representation pair, \ie $P\in \{(r,d),(d,r),(c,d),(d,c),(b,d),(d,b)\}$.

\begin{figure}[h]
\centering
\scriptsize
\begin{minipage}[c]{.85\linewidth}
\begin{minipage}[c]{.25\linewidth}
\includegraphics[width = .95\linewidth]{matc}
 \centerline{(a)}
\end{minipage}\hfill
\begin{minipage}[c]{.25\linewidth}
\includegraphics[width = .95\linewidth]{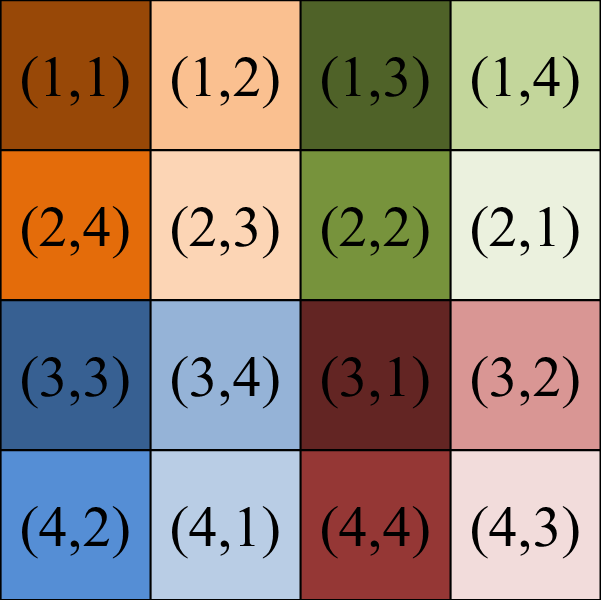}
 \centerline{(c)}
\end{minipage}\hfill
\begin{minipage}[c]{.25\linewidth}
\includegraphics[width = .95\linewidth]{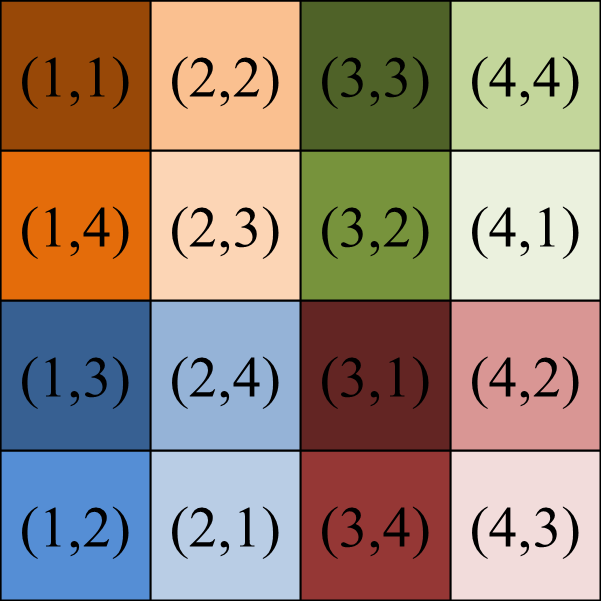}
 \centerline{(e)}
 \end{minipage}\hfill
 \begin{minipage}[c]{.25\linewidth}
\includegraphics[width = .95\linewidth]{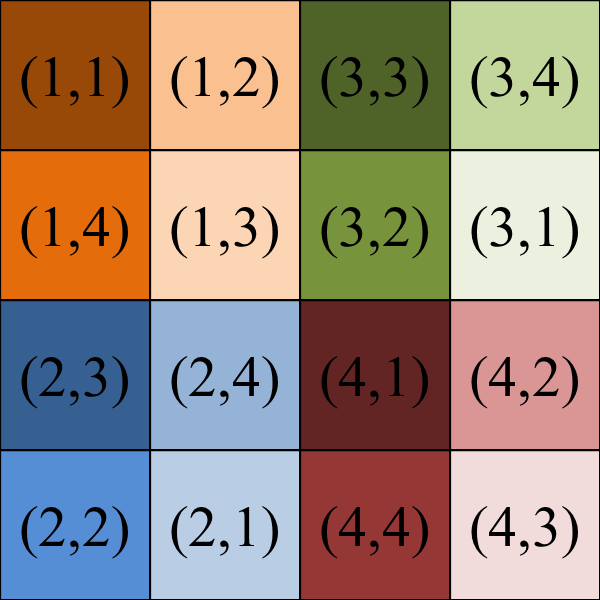}
 \centerline{(g)}
 \end{minipage}\hfill
 \begin{minipage}[c]{.25\linewidth}
\includegraphics[width = .95\linewidth]{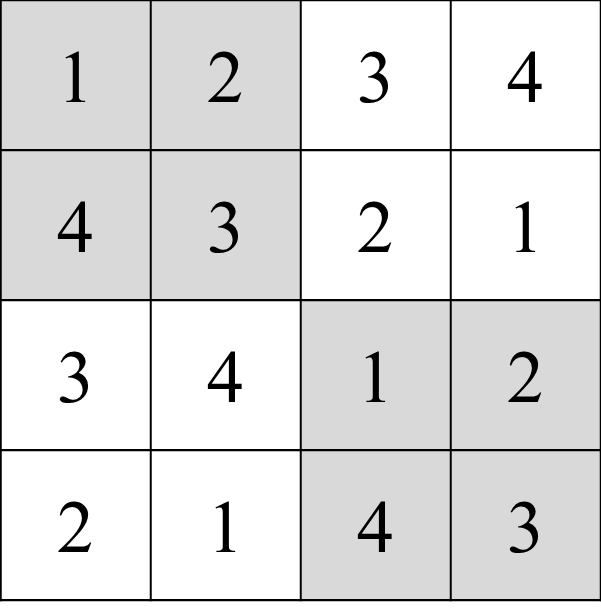}
 \centerline{(b)}
\end{minipage}\hfill
\begin{minipage}[c]{.25\linewidth}
\includegraphics[width = .95\linewidth]{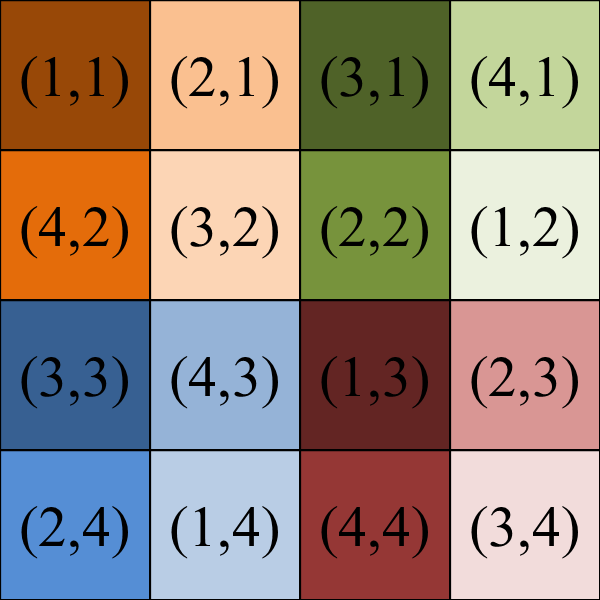}
 \centerline{(d)}
\end{minipage}\hfill
\begin{minipage}[c]{.25\linewidth}
\includegraphics[width = .95\linewidth]{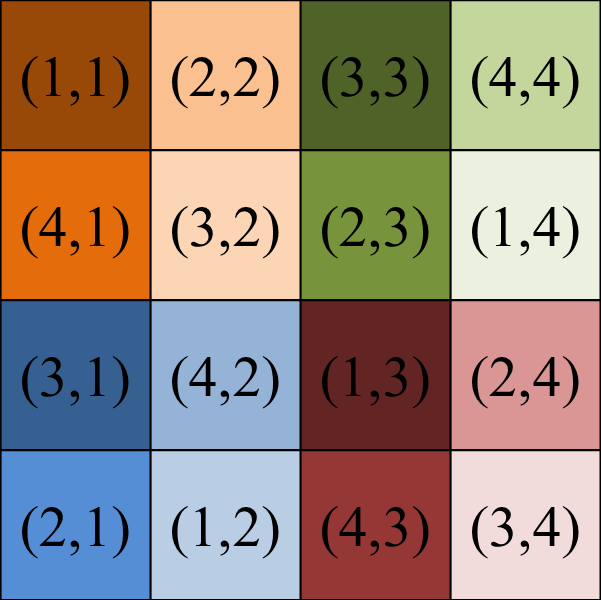}
 \centerline{(f)}
 \end{minipage}\hfill
 \begin{minipage}[c]{.25\linewidth}
\includegraphics[width = .95\linewidth]{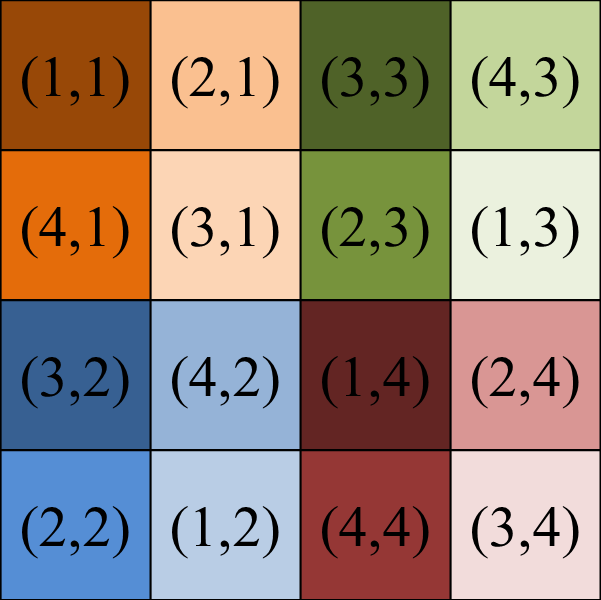}
 \centerline{(h)}
 \end{minipage}\hfill
  \end{minipage}\hfill
\caption{Sudoku associated matrix element representations; (a) a $4\times 4$ matrix $M$ with elements identified in distinctive colors;(b) reference Sudoku matrix $S$; (c) $R^S_{(r,d)}$ representation;  (d) $R^S_{(d,r)}$ representation;(e) $R^S_{(c,d)}$ representation; (f) $R^S_{(d,c)}$ representation; (g) $R^S_{(b,d)}$ representation; and (h) $R^S_{(d,b)}$ representation.}
\label{fig:SudokuNotation}
\end{figure}

Fig. \ref{fig:SudokuNotation} shows the six Sudoku associated matrix element representations for a $4\times 4$ matrix, when the reference Sudoku $S$ shown in Fig. \ref{fig:SudokuNotation}-(b) is used. It is noticeable that each representation is able to uniquely locate an element in $M$. Details about why these Sudoku associated matrix element representations are valid are given in Appendix.

\begin{figure}[h]
\centering
\scriptsize
\begin{minipage}[c]{.85\linewidth}
\begin{minipage}[c]{.25\linewidth}
\includegraphics[width = .95\linewidth]{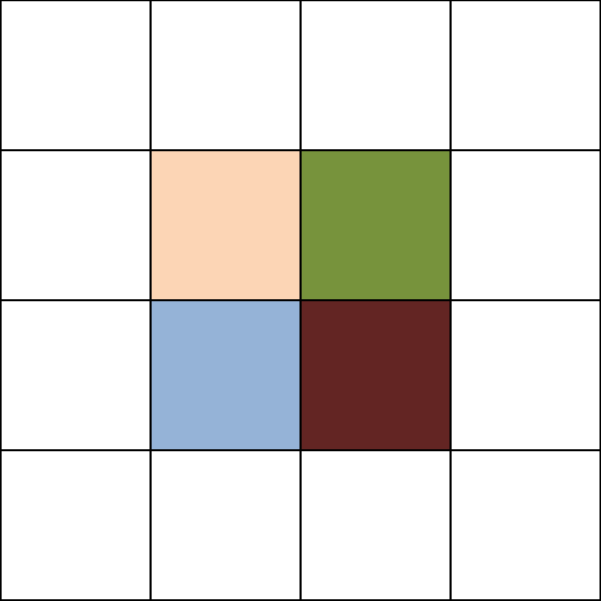}
 \centerline{(a)}
\end{minipage}\hfill
\begin{minipage}[c]{.25\linewidth}
\includegraphics[width = .95\linewidth]{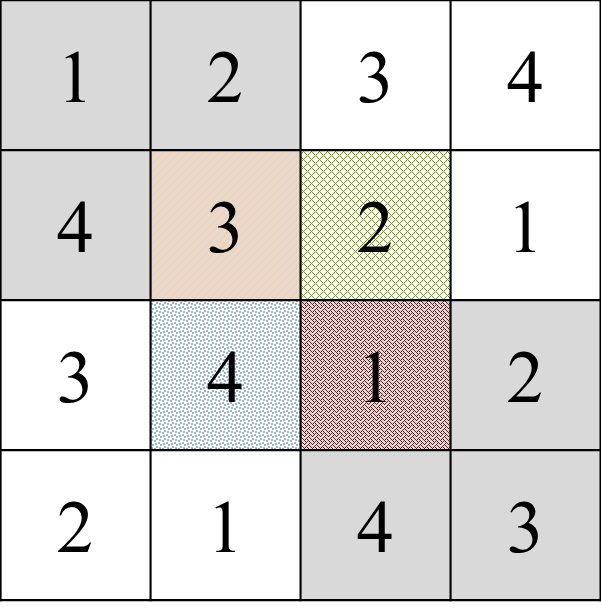}
 \centerline{(b)}
\end{minipage}\hfill
\begin{minipage}[c]{.25\linewidth}
\includegraphics[width = .95\linewidth]{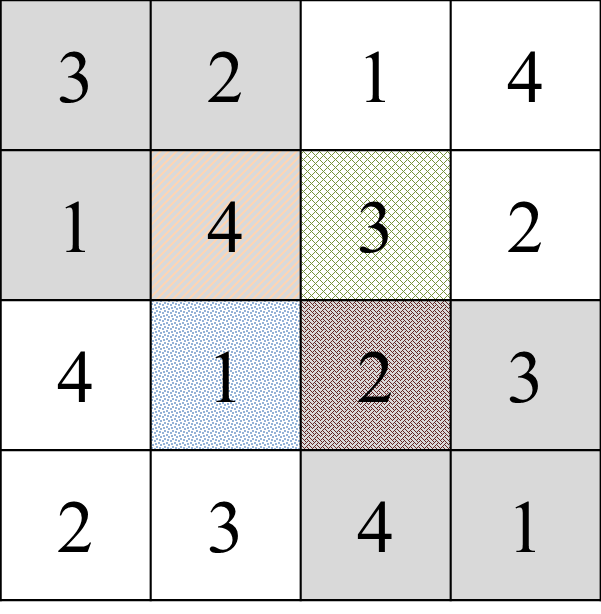}
 \centerline{(c)}
 \end{minipage}\hfill
 \begin{minipage}[c]{.25\linewidth}
\includegraphics[width = .95\linewidth]{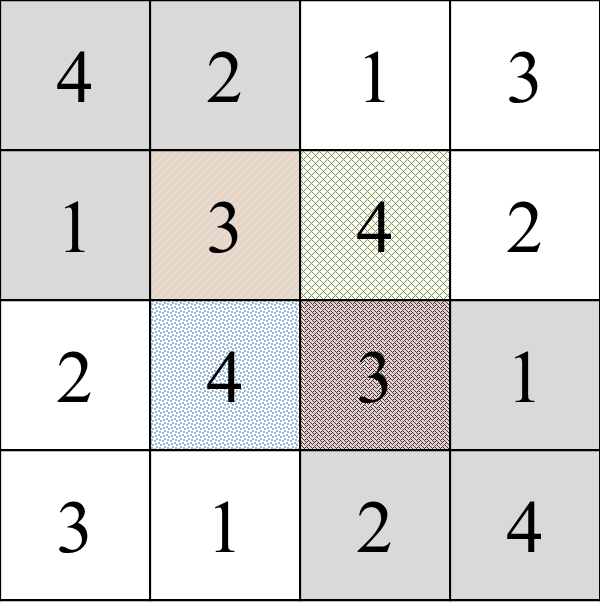}
 \centerline{(d)}
 \end{minipage}\hfill
  \end{minipage}\hfill
\caption{A $4\times 4$ matrix with colored elements and two Sudoku matrices; (a) interested matrix elements identified by colors; (b) Sudoku matrix $S_1$; (c) Sudoku matrix $S_2$ and (d) Sudoku matrix $S_3$.}
\label{fig:referenceMat}
\end{figure}

Besides the existence of the six Sudoku associated matrix element representations, we also want to emphasize that for the same Sudoku associated matrix element representation, it could be very different when the reference Sudoku matrices are different. Here is a simple example illustrating this fact for interested elements in a $4\times 4$ matrix. Fig. \ref{fig:referenceMat}-(a) shows our interested elements, and -(b), (c), (d) give three reference Sudoku matrices. Then for each interested element, we are then able to represent it with various Sudoku associated matrix element representations. For example, the \textit{green} element located at $(2,3)$ in the conventional row-column representation is denoted as $(2,2)$ by using the $(r,d)$ representation pair associated with Sudoku $S_1$, because the element located at row index $r=2$ and digit index $d=2$ in $S_1$ is the green element. In the same manner, this green element can be also denoted as $(4,3)$ by using the representation $R_{(d,b)}$ associated with Sudoku $S_3$. And Table \ref{tab:referenceMat} shows the six matrix element representations for interested elements associated reference Sudoku matrix $S_1, S_2$ and $S_3$. As can be seen from this table, a matrix element can be represented with Sudoku associated representation pairs including $(r,d), (d,r), (c,d), (d,c), (b,d)$ and $(d,b)$; and a matrix element can be represented differently by one Sudoku associated matrix element representation, if its associated reference Sudoku matrix changes. The reason behind such differences is that different Sudoku matrices have different digits in a grid. Moreover, due to the fact that these Sudoku associated matrix element representations are sensitive to the reference Sudoku matrix, we are therefore able to construct parametric two dimensional bijective functions for image scrambling.

\begin{table}[h]
  \caption{Matrix Element Representations Associated with a Reference Sudoku Matrix}\label{tab:referenceMat}
\scriptsize
\centering
\begin{tabular}{r|cccc|cccc|cccc}
\hline\hline
&\multicolumn{4}{c|}{\textbf{$S_1$}}&\multicolumn{4}{c|}{\textbf{$S_2$}}&\multicolumn{4}{c}{\textbf{$S_3$}}\\
\textbf{Representation Pair} & \textbf{\color{color1}{$\blacksquare$}} & \textbf{\color{color2}{$\blacksquare$}}& \textbf{\color{color3}{$\blacksquare$}} &\textbf{\color{color4}{$\blacksquare$}}  & \textbf{\color{color1}{$\blacksquare$}} & \textbf{\color{color2}{$\blacksquare$}}& \textbf{\color{color3}{$\blacksquare$}} &\textbf{\color{color4}{$\blacksquare$}} & \textbf{\color{color1}{$\blacksquare$}} & \textbf{\color{color2}{$\blacksquare$}}& \textbf{\color{color3}{$\blacksquare$}} &\textbf{\color{color4}{$\blacksquare$}}\\\hline
$(r,d)$ & $(2,3)$ & $(2,2)$ & $(3,4)$ &$(3,1)$ & $(2,4)$ & $(2,3)$ & $(3,1)$ &$(3,2)$ & $(2,3)$ & $(2,4)$ & $(3,4)$ &$(3,3)$\\
$(d,r)$ & $(3,2)$ & $(2,2)$ & $(4,3)$ &$(1,3)$ & $(4,2)$ & $(3,2)$ & $(1,3)$ &$(2,3)$ & $(3,2)$ & $(4,2)$ & $(4,3)$ &$(3,3)$\\
$(c,d)$ & $(2,3)$ & $(3,2)$ & $(2,4)$ &$(3,1)$ & $(2,4)$ & $(3,3)$ & $(2,1)$ &$(3,2)$ & $(2,3)$ & $(3,4)$ & $(2,4)$ &$(3,3)$\\
$(d,c)$ & $(3,2)$ & $(2,3)$ & $(4,2)$ &$(1,3)$ & $(4,2)$ & $(3,3)$ & $(1,2)$ &$(2,3)$ & $(3,2)$ & $(4,3)$ & $(4,2)$ &$(3,3)$\\
$(b,d)$ & $(1,3)$ & $(3,2)$ & $(2,4)$ &$(4,1)$ & $(1,4)$ & $(3,3)$ & $(2,1)$ &$(4,2)$ & $(1,3)$ & $(3,4)$ & $(2,4)$ &$(4,3)$\\
$(d,b)$ & $(3,1)$ & $(2,3)$ & $(4,2)$ &$(1,4)$ & $(4,1)$ & $(3,3)$ & $(1,2)$ &$(2,4)$ & $(3,1)$ & $(4,3)$ & $(4,2)$ &$(3,4)$\\\hline\hline
\end{tabular}
\end{table}

It is clear that for each Sudoku matrix, there are six associated matrix element representations. Since any two matrix element representation denote the same set of matrix elements, a bijective mapping then can be constructed correspondingly like we showed in Fig. \ref{fig:rcbk}-(d). Since we have two fixed matrix element representations, namely row-column pair and block-grid pair, we are able to construct two new bijections denoted as $f_{R^S_P\rightarrow R_F}$ by mapping from the Sudoku associated matrix element representation $R^S_P$ to the two fixed representations $R_F$, and another two new bijections by mapping from the two fixed representations to the Sudoku associated matrix element representation denoted as $f_{R^S_P\leftarrow R_F}$, where $S$ denotes the reference Sudoku matrix, $R$ denotes a matrix element representation, $P$ denotes a Sudoku associated representation pair with $P\in \{(r,d),(d,r),(c,d),(d,c),(b,d),(d,b)\}$, and $F$ denotes a fixed matrix element representation with $F\in \{(r,c),(b,g)\}$. In summary, the total number Sudoku associated bijections can be directly constructed between a Sudoku associated matrix element representation to a fixed element representation is $6\times 2\times 2 = 24$.

\begin{figure}[h]
\centering
\scriptsize
\begin{minipage}[c]{.95\linewidth}
\centering
\begin{minipage}[c]{.5\linewidth}
 \centerline{\includegraphics[width = .5\linewidth]{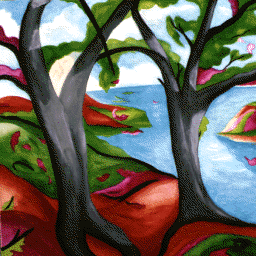}}
 \centerline{(a)}
\end{minipage}\hfill
\begin{minipage}[c]{.5\linewidth}
 \centerline{\includegraphics[width = .5\linewidth]{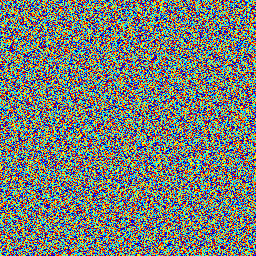}}
 \centerline{(b)}
\end{minipage}\hfill
\end{minipage}\hfill
\begin{minipage}[c]{1\linewidth}
\begin{minipage}[c]{1\linewidth}
 \centerline{\includegraphics[width = 1\linewidth]{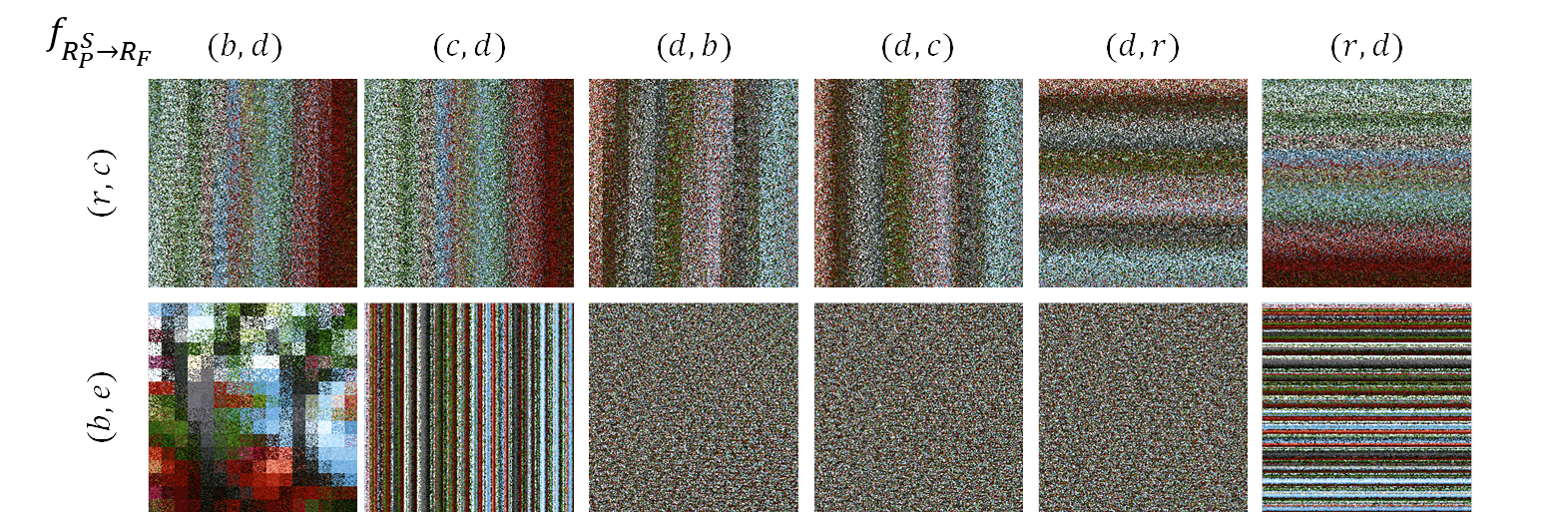}}
 \centerline{(c)}
\end{minipage}\hfill
\begin{minipage}[c]{1\linewidth}
 \centerline{\includegraphics[width = 1\linewidth]{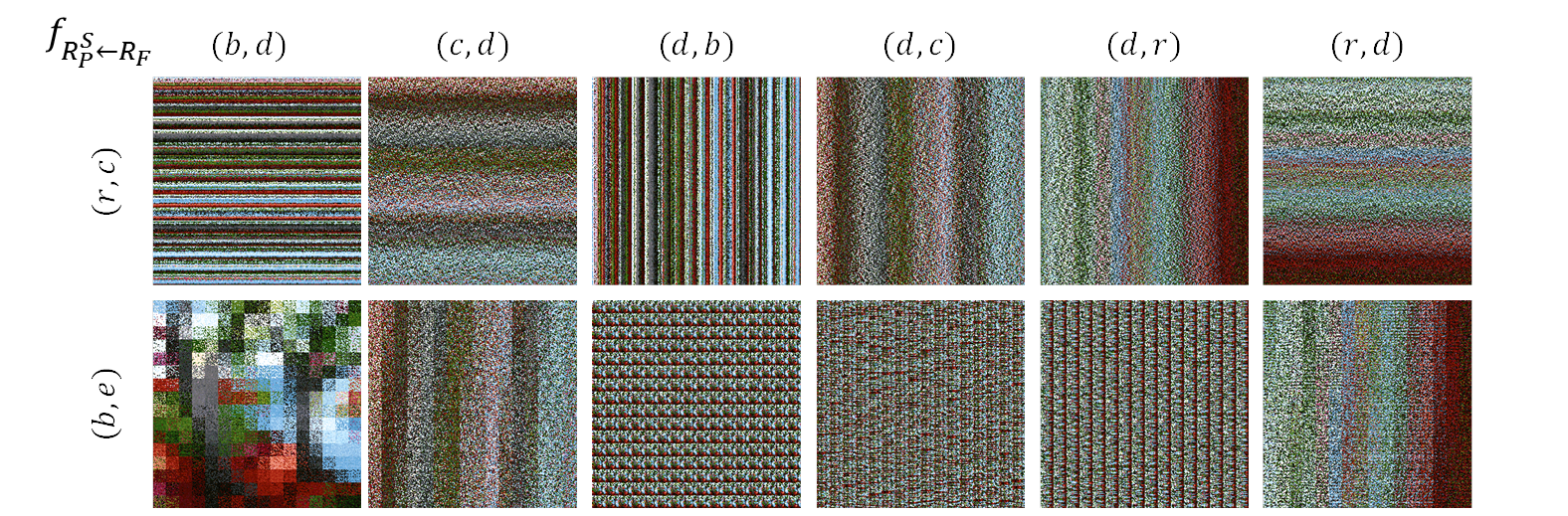}}
 \centerline{(d)}
 \end{minipage}\hfill
 \end{minipage}\hfill
\caption{One-step image scrambling results using the Sudoku associated bijections (a) input image $256\times 256$ \textit{Trees}; (b) a $256\times 256$ Sudoku matrix $S$; (c) scrambling results with the twelve bijections $f_{R^S_P\rightarrow R_F}$; and (d) scrambling results with the twelve bijections $f_{R^S_P\leftarrow R_F}$. }
\label{fig:map48}
\end{figure}

Fig. \ref{fig:map48} shows a naive one step image scrambling using the twenty-four Sudoku associated bijections. As can be seen from these results, different Sudoku associated bijections scramble the original \textit{Trees} image differently. For example, the bijections $f_{R^S_{(r,d)}\rightarrow R_{(r,c)}}$ and $f_{R^S_{(r,c)}\leftarrow R_{(r,d)}}$ shuffle image pixels within each row; the bijections $f_{R^S_{(d,c)}\rightarrow R_{(r,c)}}$ and $f_{R^S_{(d,c)}\leftarrow R_{(r,c)}}$ shuffle image pixels within each column; the bijections $f_{R^S_{(d,b)}\rightarrow R_{(b,g)}}$ and $f_{R^S_{(d,b)}\leftarrow R_{(b,g)}}$ shuffle image pixels within each $16\times 16$ block. It is also noticeable that $f_{R^S_{(d,b)}\rightarrow R_{(b,g)}}$, $f_{R^S_{(d,c)}\rightarrow R_{(b,g)}}$, and $f_{R^S_{(d,r)}\rightarrow R_{(b,g)}}$ are the three most powerful bijections that scramble the original \textit{Trees} image to almost random-like from the point view of human vision inspection.

\subsection{More Sudoku Associated Bijections}
It is well known that the composition of two bijective mapping is still a bijective mapping. In other words, if both $f_0$ and $f_1$ are two bijection on $\mathbb{I}\times \mathbb{I}$, then $f_{new} = f_0\circ f_1$ is also a bijection on $\mathbb{I}\times \mathbb{I}$ and so is $f_{new} = f_1\circ f_0$ . Therefore, we can even find more two dimensional bijective mappings by simply composing two or more existent bijective mappings.

\begin{table}[h]
  \caption{The number of two dimensional bijective mappings associated with Sudoku matrices}\label{tab:bijectionNum}
\scriptsize
\centering
\begin{tabular}{r|rrrr}
\hline\hline
{\textbf{\# of two dimensional bijection: ${{\cal{N}}^{t}_s}$}}&\multicolumn{4}{c}{\textbf{\# of Sudoku matrices: $s$}} \\\cline{1-1}
{\textbf{Composition times $t$}}  & \textbf{1} & \textbf{2} &$\mathbf{\cdots}$ & $\mathbf{m}$\\\cline{2-5}
\textbf{$t = 0$}& 24 & 48  &$\cdots$ & $24m$\\
\textbf{$t = 1$}& $24^2$ & $48^2$ & $\cdots$ & $(24m)^2$\\
\textbf{$\vdots$}&  $\vdots$& $\vdots$ &$\ddots$ & $\vdots$\\
\textbf{$t = j$} & $24^{j+1}$ & $48^{j+1}$ &$\cdots$ &$(24m)^{j+1}$\\
 \hline\hline
\end{tabular}
\end{table}

Table \ref{tab:bijectionNum} shows the relations of the number of reference Sudoku matrices and the number of two dimensional bijective mappings. When one Sudoku matrix is used for reference, then six new matrix element representations can be found. Therefore, when $m$ Sudoku matrices are used, there are $6m$ Sudoku associated matrix element representations. Since a bijection can be constructed between a Sudoku associated matrix element representation and one fixed representation, the number of bijections available is $2\times 2\times 6m = 24m$, where $6m$ is the number of Sudoku associated matrix element representations, $2$ is the number of fixe matrix element representations, and $2$ implies the two possible ways to construct a bijection either mapping from the fixed representation to the Sudoku associated one or from the Sudoku associated one to the fixed one.

Because the composition of two bijections is just a new bijection, we can define more bijections by involving function compositions. Without loss of generality, say we are interested in the number of bijections defined by $t$ times of function compositions from $i$ bijections $f_1,f_2,\cdots,f_i$, \ie
$$f_{i_0\circ i_1\cdots\circ i_t} = \underbrace{f_{i_0}\circ f_{i_1}\cdots\circ f_{i_t}}_{t \textrm{ times}}$$
where $i_0,i_1,\cdots,i_t \in \{1,2,\cdots, i\}$. Then it is not difficult to see that the total number of bijections by $t$ times of compositions is $i^{t+1}$, since for each function $f_{i_j}$ with $j\in \{0,1,\cdots,t\}$ we have $i$ candidate function and in total we have $t+1$ functions to be determined. In Table \ref{tab:bijectionNum} we have $24m$ of possible Sudoku associated matrix element representations and $j$ times of compositions \ie $i=24m$ and $t = j$, and thus the number of bijections involving $j$ times of compositions is $i^{t+1} = (24m)^{j+1}$.

It is worthwhile to note the number of bijections calculated in Table \ref{tab:bijectionNum} includes self-mapping, \ie $f_{new} = f_0\circ f_1$ with $f_1 = f_0^{-1}$. However, as long as the number of Sudoku matrices increases the possibility of randomly composing a pair of bijections (one is the inverse of the other) quickly approach to zero. An alternative remedy to avoid self-mapping is to choose the two Sudoku associated bijections with different reference Sudoku matrices, so that a new composed bijection from these two bijections is impossible to be a self-mapping unless the two reference Sudoku matrices are identical.

\section{Sudoku Associated Image Scrambling Scheme}
In previous sections, we showed that given a $N\times N$ Sudoku matrix $S$ for reference, there are six Sudoku associated matrix element representations, namely the row-digit pair representation $R^S_{(r,d)}$, digit-row pair representation $R^S_{(d,r)}$, column-digit pair representation $R^S_{(c,d)}$, digit-column pair representation $R^S_{(d,c)}$, block-digit pair representation $R^S_{(b,d)}$ and digit-block pair representation $R^S_{(d,b)}$. Each Sudoku $S$ associated matrix element representation $R^S_P$ with $P\in \{(r,d), (d,r), (c,d), (d,c), (b,d),(d,b)\}$ and a fixed representation $R_F$ with $F\in \{(r,c),(b,g)\}$ can be used to construct a pair of bijections $f_{R^S_P\leftarrow R_F}$ and $f_{R^S_P\rightarrow R_F}$, where $f_{R^S_P\leftarrow R_F}$ denotes the bijection mapping from the fixed representation $R_F$ to the representation $R^S_P$, and $f_{R^S_P\rightarrow R_F}$  denotes the bijection mapping from the Sudoku associated representation $R^S_P$ to the fixed representation $R_F$. Therefore, given a Sudoku matrix $S$ and a fixed representation $R_F$, there are 24 Sudoku associated bijective mappings of $f_{R\rightarrow F}^{S}$ and $f_{R\leftarrow F}^{S}$. In this section, we use these two dimensional Sudoku associated bijections for digital image scrambling.

\subsection{Sudoku Associated Image Scrambler}
\begin{figure}[h]
\scriptsize
\centering
  \includegraphics[width=.85\linewidth]{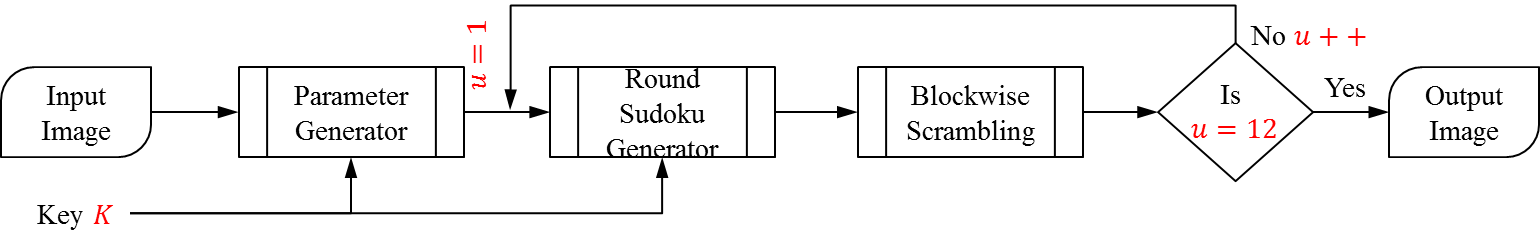}\\
  \centerline{(a)}
    \includegraphics[width=.85\linewidth]{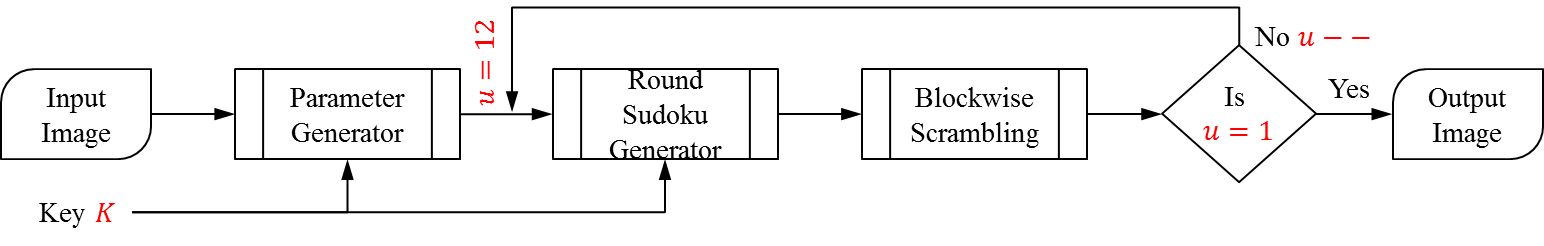}\\
     \centerline{(b)}
  \caption{Flowchart of a Sudoku associated image scrambler; (a) scrambling process; (b) descrambling process}\label{fig:scrambler}
\end{figure}

Although a Sudoku associated image scrambler can be designed in various means, we design a Sudoku associated image scrambler as shown in Fig. \ref{fig:scrambler}, where \textit{Key} is an encryption key of 192-bit length, \textit{Parameter Generator} generates key dependent and round dependent parameters including a Sudoku associated two dimensional bijection $f_{R^S_P\rightarrow R_F}$ or $f_{R^S_P\leftarrow R_F}$, \textit{Blockwise Scrambling} shuffles pixels within a $N\times N$ image block for every image block of the input image. Consequently, the image descrambling process is just the reverse of the scrambling process as shown in Fig. \ref{fig:scrambler}-(b). In the following discussion, we consider an input image $X$  is of size $W\times H$ with $nB$ bit-depth.

\textit{Parameter Generator} translates a scrambling key $K$ to required parameters for image scrambling. Specifically, it takes a scrambling key $K$ to generate a set of parameters including twelve Sudoku associated matrix element representation pairs $\mathbf{P^i} = \{P_1^i,P_2^i,\cdots,P_{12}^i\}$, twelve fixed representation pair $\mathbf{F^i} = \{F_1^i,F_2^i,\cdots,F_{12}^i\}$ and the twelve mapping parameters $\mathbf{m^i} = \{m_1^i,m_2^i,\cdots,m_{12}^i\}$ for the $i$th bit-plane in $X$ as shown in Algorithm 1. In this way, \textit{Parameter Generator} guarantees that for each bit-plane in $X$ all six possible Sudoku associated matrix element representation pairs and two possible fixed representation pairs are used in the twelve round scrambling. Meanwhile, these parameters are bit-plane dependent, implying different bit-planes will be scrambled in a different way in the future processing.

\begin{algorithm}
\caption{\textbf{Parameter Generator $(\mathbf{R^i},\mathbf{F^i},\mathbf{m^i}) = BPG(K,i)$}}
\scriptsize
\begin{algorithmic}
\REQUIRE $K$ is a key of 192 bits composed of twenty-four subkeys $\{K^{(1)},K^{(2)},\cdots,K^{(24)}\}$
\REQUIRE $i$ is the $i$th bit-plane to be processed
\ENSURE $(\mathbf{P},\mathbf{F},\mathbf{m})$ is a list of twelve pairs of Sudoku associated matrix element representation pairs, fixed matrix element representation pairs, and mapping directions.
\STATE $[K^{(1)},K^{(2)},\cdots,K^{(24)}] = K$ \% Divide key $K$ to twenty-four subkeys $K^{(1)},K^{(2)},\cdots,K^{(24)}$
\STATE $Q = [K^{i},K^{mod(i+1,24)+1},\cdots, K^{mod(i+12,24)+1}]$ \% Extract a twelve element subkey sequence starting at $K^{(i)}$
\STATE $idx = sort(Q)$ \% Sort this sequence $Q$ and generate the element index sequence in the sorted sequence
\FOR{$u = 1:1:12$}
    \STATE $q = idx(j)$ \% For each index in $idx$, pair it to a $(R,F)$ representations
    \IF {$mod(q,2) == 1$}
    \STATE $m^i_u = 1$
    \ELSE
    \STATE $m^i_u = 0$
    \ENDIF
    \IF {$q == 1$}
    \STATE $P_u^i = (b,d), F_u^i = (r,c)$
    \ENDIF
    \IF {$q == 2$}
    \STATE $P_u^i = (b,d), F_u^i = (b,g)$
    \ENDIF
    \IF {$q == 3$}
    \STATE $P_u^i = (c,d), F_u^i = (r,c)$
    \ENDIF
    \IF {$q == 4$}
    \STATE $P_u^i = (c,d), F_u^i = (b,g)$
    \ENDIF
    \IF {$q == 5$}
    \STATE $P_u^i = (d,b), F_u^i = (r,c)$
    \ENDIF
    \IF {$q == 6$}
    \STATE $P_u^i = (d,b), F_u^i = (b,g)$
    \ENDIF
    \IF {$q == 7$}
    \STATE $P_u^i = (d,c), F_u^i= (r,c)$
    \ENDIF
    \IF {$q == 8$}
    \STATE $P_u^i = (d,c), F_u^i = (b,g)$
    \ENDIF
    \IF {$q == 9$}
    \STATE $P_u^i = (d,r), F_u^i = (r,c)$
    \ENDIF
    \IF {$q == 10$}
    \STATE $P_u^i = (d,r), F_u^i = (b,g)$
    \ENDIF
    \IF {$q == 11$}
    \STATE $P_u^i = (r,d), F_u^i = (r,c)$
    \ENDIF
    \IF {$q == 12$}
    \STATE $P_u^i = (r,d), F_u^i = (b,g)$
    \ENDIF
\ENDFOR
\end{algorithmic}
\end{algorithm}

Furthermore, \textit{Round Sudoku Generator} takes the image size and the scrambler round $u$ to generate a $N\times N$ Sudoku matrix as shown in Algorithm 2.
\begin{algorithm}
\caption{\textbf{Round Sudoku Generator $ S_u = RSG(K,u,W,H)$}}
\scriptsize
\begin{algorithmic}
\REQUIRE $K$ is a key of 256 bits
\REQUIRE $u$ is the cipher round number
\ENSURE $S_u$ is a $N\times N$ Sudoku matrix
\STATE $K_u = RoundKeyGenerator(K,u)$ \% Generate a round key \footnotemark[2]
\STATE $N= \lfloor\sqrt{\min(W,H)}\rfloor$ \% Determine Sudoku size
\STATE $S_u = SudokuGenerator(K_u,N)$ \% Generate a $N\times N$ Sudoku matrix
\end{algorithmic}
\end{algorithm}

Consequently, a two dimensional Sudoku associated bijection $f_u^i$ can be found for the $i$th bit-plane in $u$th scrambler round as follows
\begin{equation}\label{eq:mapping}
    f_u^i = \left\{\begin{array}{l}
        f_{R^{S_u}_{P^i_u}\leftarrow F^i_u}\textrm{, if } m^i_u = 0\\
        f_{R^{S_u}_{P^i_u}\rightarrow F^i_u}\textrm{, if } m^i_u = 1
    \end{array}\right.
\end{equation}

Once the $N\times N$ parametric Sudoku associated two dimensional bijection $f^i_u$ is obtained for the $i$th bit-plane in the $u$th scrambler round, we are then able to scramble image pixels with in $N\times N$ image within the $W\times H$ input image for all of its bit-planes. In order to obtain pixel shuffling between blocks, we apply a image shift function ${imgShift(.)}$ for each cipher round as shown in Eq. \eqref{eq:imgShift}.
\begin{equation}\label{eq:imgShift}
    Y = imgShift(X,w,h) = \left[\begin{array}{c|c}
    X(w:W,h:H)& X(w:W,1:h-1)\\\hline
    X(1:w-1,h:H) & X(1:w-1,1:h-1)
    \end{array}\right]
\end{equation}
Pseudo-code of \textit{Blockwise Scarmbling} algorithm is given below in the MATLAB fashion. Consequently, we then shuffled every $N\times N$ image blocks within the input image $X$.
\begin{algorithm}
\caption{\textbf{Blockwise Scrambling $Y = BlkScarmbling(X,\mathbf{f_u},N,nB)$}}
\scriptsize
\begin{algorithmic}
\REQUIRE $I$ is an image of size $W\times H$
\REQUIRE $\mathbf{f_u} = \{f_u^1,f_u^2,\cdots,f_u^{nB}\}$ is a vector of $nB$ bijections with each bijection for one bit-plane
\REQUIRE $N$ is a squared integer
\REQUIRE $nB$ is the number of bit-planes contained in image $X$
\ENSURE $Y$ is an scrambled image of $X$ at the same size
\STATE $nRow = \lceil{W/N}\rceil$
\STATE $nCol = \lceil{H/N}\rceil$
\STATE $w = \lceil{W/12}\rceil$
\STATE $h = \lceil{H/12}\rceil$
\STATE $X = imgShift(X,w,h)$ \% shift image $X$ $w$ pixels along rows and $h$ pixels along columns
\FOR{$i = 1:1:nRow$}
    \IF{$i~=nRow$}
        \STATE $r0 = (i-1)N+1$ \% 1st row index of the current block
        \STATE $r1 = iN$ \% last row index of the current block
    \ELSE
        \STATE $r0 = W-N+1$ \% 1st row index of the current block
        \STATE $r1 = W$ \% last row index of the current block
    \ENDIF
    \FOR{$j = 1:1:nCol$}
        \IF{$j~=nCol$}
            \STATE $c0 = (j-1)N+1$ \% 1st column index of the current block
            \STATE $c1 = jN$ \% last column index of the current block
        \ELSE
            \STATE $c0 = H-N+1$ \% 1st column index of the current block
            \STATE $c1 = H$ \% last column index of the current block
        \ENDIF
        \STATE{$tBlk = I(r0:r1,c0:c1)$} \% extract the current block
        \STATE{$tmp = 0$} \% create a temporary intermediate variable
        \FOR{$l = 1:1:nB$}
            \STATE $tBP = bitget(tBlk,l)$ \% extract the $u$th bit-plane
            \STATE{$oBP(r0:r1,c0:c1) = f^l_u(tBP)$} \% output scrambled bit-plane
            \STATE $tmp = tmp+oBP\cdot 2^{l-1}$ \% store results in temporary variable
        \ENDFOR
        \STATE $Y(r0:r1,c0:c1) = tmp$ \% output scrambled block
    \ENDFOR
\ENDFOR
\end{algorithmic}
\end{algorithm}

\begin{figure}[h]
\centering
\scriptsize
  \includegraphics[width=.5\linewidth]{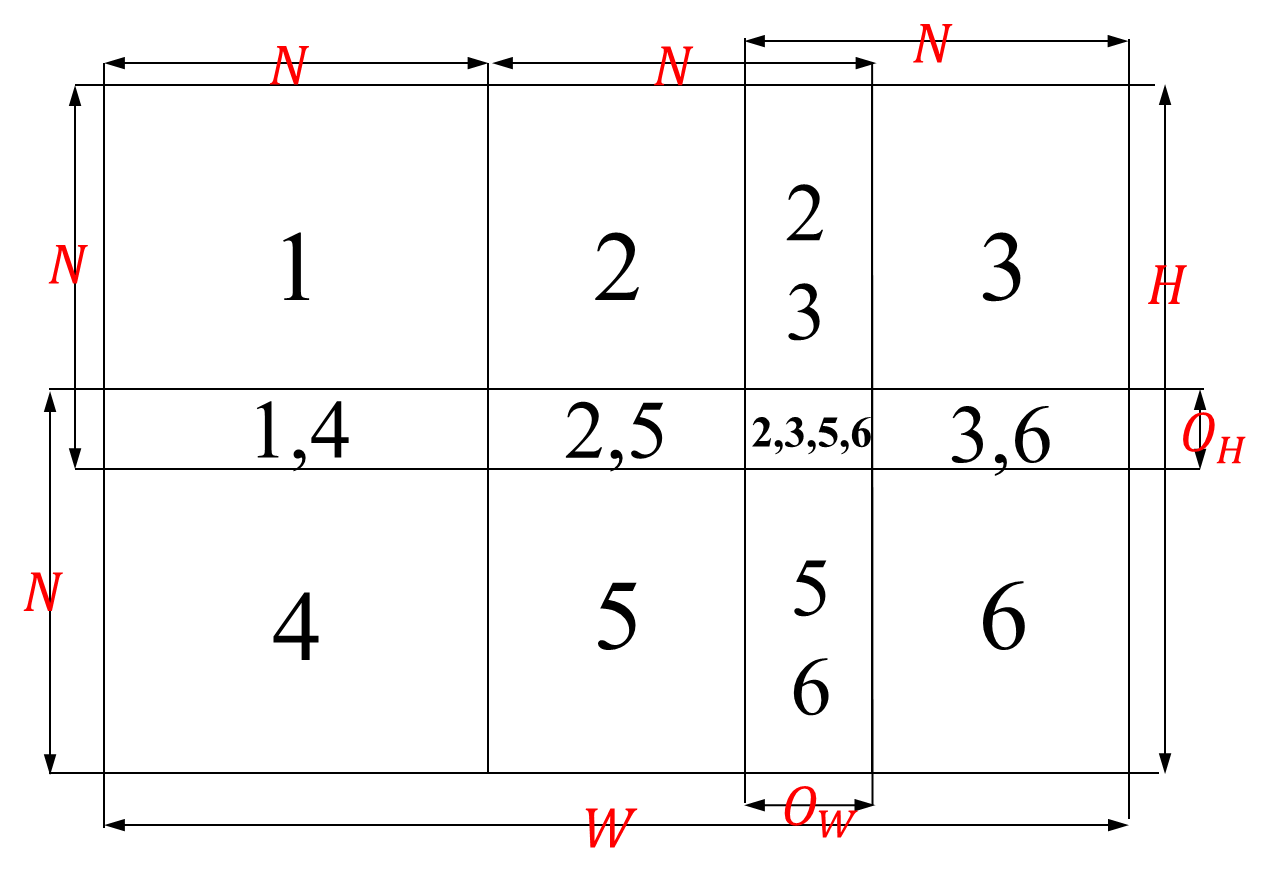}\\
  \caption{Overlapped blocks in \textit{Blockwise Scrambling}}\label{fig:overlapped}
\end{figure}

In order to scramble an arbitrary $W\times H$ image without bandwidth expansion, we scramble blocks on the edge with a certain amount of overlapping with previous blocks as shown in Fig. \ref{fig:overlapped}. The block indices listed in Fig. \ref{fig:overlapped} indicate the processing order of \textit{Blockwise Scrambling}, implying to first shuffle image pixels in the upper-left block marked as index 1 in the figure, then block 2, block 3, so on and so forth until block 6. Since $W$ and $H$ may not be necessarily to be divisible by $N$, we process blocks on the image edge with overlapping, \eg block 2 and 3 are overlapped, and block 3 and block 6 are also overlapped. In summary, there are $O_H\times O_W$ pixels belongs to more than one block to be processed. In this way, we are able to fit a two dimensional Sudoku associated bijection defined on $\mathbb{I}\times \mathbb{I}$ for an input image with an arbitrary size $W\times H$ without bandwidth expansion.

In regarding of descrambling, we can simply substitute a certain Sudoku associated bijection with its inverse. In other words, if the bijection used for scrambling is $f_{R^S_P\leftarrow R_F}$, then the bijection used for descrambling is $f_{R^S_P\rightarrow R_F}$, and vice versa.
\subsection{Discussions}
\textit{Sudoku Associated Image Scrambler} is a multiround scrambler that shuffles image pixels with respect to one Sudoku associated bijection for each round. This multiround scrambling process is equivalent to create a new bijection by taking the function composition of multiple bijections, which is discussed in Section III-C. It is not difficult to validate such equivalence between the multiround Sudoku associated image scrambler and the bijection by composing multiple bijections. Without loss of generality, denote the twelve two dimensional Sudoku associated bijections for the $i$th bit-plane are $f_1^i,f_2^i,\cdots,f_{12}^i$ in \textit{Sudoku Associated Image Scrambler}. Then running the scrambler for multiple times is nothing but to construct a new bijection $f_{new}$ as follows
\begin{equation}\label{eq:fnew}
    f_{new}(x,y) = f_{12}^i(f_{11}^i(\cdots f_{2}^i(f_1^i(x,y)))) = f_{12}^i\circ f_{11}^i\cdots\circ f_{2}^i\circ f_{1}^i(x,y)
\end{equation}
which is actually the composition of the twelve bijections.

\begin{figure}[h]
\centering
\scriptsize
\begin{minipage}[c]{.95\linewidth}
\begin{minipage}[c]{.2\linewidth}
\includegraphics[width = .95\linewidth]{Trees}
 \centerline{(a)}
\end{minipage}\hfill
\begin{minipage}[c]{.2\linewidth}
\includegraphics[width = .95\linewidth]{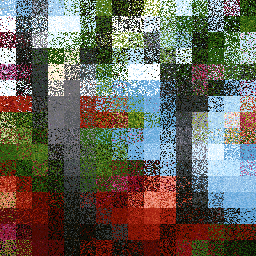}
 \centerline{(b)}
\end{minipage}\hfill
\begin{minipage}[c]{.2\linewidth}
\includegraphics[width = .95\linewidth]{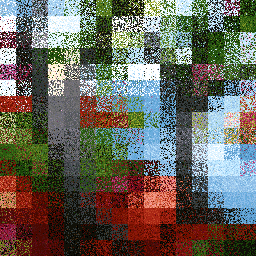}
 \centerline{(c)}
 \end{minipage}\hfill
 \begin{minipage}[c]{.2\linewidth}
\includegraphics[width = .95\linewidth]{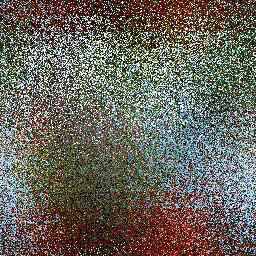}
 \centerline{(d)}
 \end{minipage}\hfill
  \begin{minipage}[c]{.2\linewidth}
\includegraphics[width = .95\linewidth]{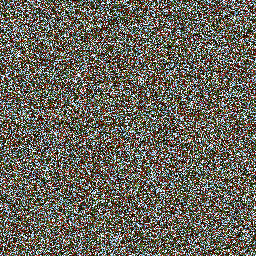}
 \centerline{(e)}
 \end{minipage}\hfill
  \end{minipage}\hfill
\caption{Image scrambling results discussion; (a) image \textit{Trees}; (b) scrambled image after one round $f_{R^S_{(b,d)}\rightarrow R_{(b,g)}}$; (c) scrambled image after twelve rounds $f_{R^S_{(b,d)}\rightarrow R_{(b,g)}}$; (d) scrambled image after twelve rounds $f_{R^S_{(b,d)}\rightarrow R_{(b,g)}}$ with image shifting; and (e) scrambled image after twelve rounds scrambling in \textit{Sudoku Associated Image Scrambler}. }
\label{fig:discussion}
\end{figure}

Meanwhile, we want to emphasize that the Sudoku associated bijections used in \textit{Sudoku Associated Image Scrambler} from one round to another is not independent, but dependent in the sense that in the twelve round the fixed representation $R_{(r,c)}$ and $R_{(b,g)}$ appears exact six times respectively and the six Sudoku associated representations appears exact twice for each. We set up this constraint to guarantee the strong scrambling effect. Although different Sudoku associate bijections have different scrambling impact on an image, many of them are of limited scrambling effect in the sense that scrambled images are not random-like as shown in Fig. \ref{fig:map48}. Therefore, if we allow random Sudoku associated bijections for each round, it is possible to pick twelve identical Sudoku associated matrix element representation pairs, \ie $P_1= P_2= \cdots = P_{12}$ and also twelve identical fixed representation pairs $F_1=F_2\cdots=F_{12}$. Although the reference Sudoku changes for each round, the scrambled image might not change that much. Fig. \ref{fig:discussion} shows an example of using bijection $f_{R^S_{(b,d)}\rightarrow R_{(b,g)}}$ for image scrambling. As this mapping implies, it only scrambles pixels within a Sudoku block. And thus scrambled image gets blurry soon after applying this bijection once. Consequently, the scrambled image after applying this bijection twelve times shown in Fig. \ref{fig:discussion}-(c) is more or less the same as the previous, indicating that simply composing a number of Sudoku bijections does not necessarily improves scrambling quality. Fig. \ref{fig:discussion} -(d) shows the scrambled image with additional image shifting operation defined in Eq. \eqref{eq:imgShift}. This result shows that this operation does help improve scrambling quality by generating a more evenly distributed scrambled image. Fig. \ref{fig:discussion} -(e) shows the scrambled image using the twelve round \textit{Sudoku Associated Image Scrambler} (in this particular case we consider image \textit{Trees} to be bit-depth 1), which is completely unrecognizable and random-like.

In regarding of scrambler security, we suggest to change a scrambler key $K$ frequently. Because an image scrambler itself is not semantically secure and is vulnerable under certain attacks, for example, a simple but effective algorithm based on the chosen plaintext attack (CPA) model could be described as follows,
\begin{enumerate}
  \item Construct a all-zero image $X$ of size $W\times H$ except the pixel $X_{i,j}\neq 0$
  \item Scramble this image $X$ by using an image scrambler and a unknown key $K$ and obtain scrambled image $Y$
  \item Find out nonzero value pixel in $Y$ and record its position $(r_{idx},c_{idx})$ as the output of a bijection $f$, \ie $f(i,j) = (r_{idx},c_{idx})$, where ${idx} = (i-1)H+j$
  \item Repeat the above three steps by changing the position of the none-zero pixel in $X$, for all $i\in \{1,2,\cdots,H\}$ and $j\in \{1,2,\cdots,W\}$
\end{enumerate}
Consequently, the bijective mapping $f$ is the equivalent bijection of the used scrambler for a $W\times H$ image under the key $K$. In other words, all $W\times H$ images scrambled by the used image scrambler under key $K$ can be perfectly cracked by inverse mapping scrambled image pixels using mapping $f^{-1}$. And the complexity of the above attack is $W\times H$.

It is worthwhile to note that the above attack is a generic attack for all image scramblers. However, one remedy to this type of attacks is to change encryption key frequently for a scrambler. For example, say we use the \textit{Sudoku Associated Image Scrambler} for high-definition television (HDTV) with the format $1080p$, namely $1920\times 1080$ pixels per frame. Then an adversary need to take $1080\times 1920 = 2073600$ frames for fully crack an scrambling key $K$. We therefore set up a key change for every $518400 = 2073600/4$ frames, which is equivalent to 6 hours of HDTV programs if the frame rate is $24$ frames per second. As a result, an adversary may at most recover 25\% of pixels in a HDTV frame, which is definitely of poor visual quality.

\section{Analysis and Comparison Results}
Digital image scrambling is to rearrange image pixels in a deterministic way but with a random-like appearance. In this section we focus our effects on performance analysis and comparisons for the \textit{Sudoku Associated Image Scrambler}.

\subsection{Experiment Settings}
The following simulations are all performed under the \textit{Windows 7} operation system with the \textit{Intel Core2 CPU @2.66GHz} and $6GB$ memory. In order to make easy comparisons, we run our \textit{Sudoku Associated Image Scrambler} on test images (8-bit grayscale) whose scrambling results are widely reported and compare our results with other peer algorithms including the scrambling method based on two dimensional cellular automata proposed by Abu Dalhoum \etal \cite{Abu}, the scrambling method based on chaos map proposed by Ye \cite{ye2010image}, the scrambling method of using cellular automata proposed by Ye \etal \cite{ye2008novel}, and the scrambling method based on ASCII code of matrix element proposed by Ye \etal \cite{ye2007image}. These test images includes \textit{Cameraman}, \textit{Barbara}, \textit{Lenna}, and test images $\#157055, \#69015$ and $\#239096$ in the Berkeley image segmentation dataset\footnotemark[3] \cite{MartinFTM01}. Since the results about these test images have been previous reported, we make comparison of the proposed scrambler with peer methods by analyzing these test images.

Additionally, we also test the performance of the proposed scrambler on other image types than (8-bit grayscale) to validate that the proposed scrambler is able to deal with an image with an arbitrary bit depth, including binary images \textit{CCITT-3} and \textit{CCITT-7} from the CCITT fax image compression dataset \footnotemark[4], 16-bit grayscale knee MRI images, and color images \textit{4.2.03} and \textit{4.2.07} from the USC-SIPI miscellaneous image dataset \footnotemark[5].

Details about these test images and results with analysis and comparisons are presented in next sections.

\footnotetext[3]{This dataset is free for non-commercial research and educational purposes, available under \url{http://www.eecs.berkeley.edu/Research/Projects/CS/vision/bsds/} \astoday.}
\footnotetext[4]{The CCITT database can be found under page: \url{http://cdb.paradice-insight.us} \astoday.}
\footnotetext[5]{The USC-SIPI image database can be found on \url{http://sipi.usc.edu/database/} \astoday.}

\subsection{Simulation Results}
\begin{figure}[!h]
\centering
\scriptsize
\begin{minipage}[c]{1\linewidth}
\begin{minipage}[c]{.2\linewidth}
\centering{\includegraphics[width = .95\linewidth]{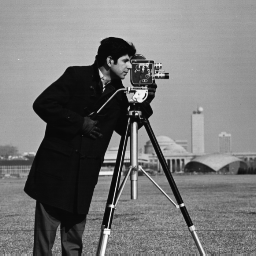}}
\end{minipage}\hfill
\begin{minipage}[c]{.2\linewidth}
\centering{\includegraphics[width = .95\linewidth]{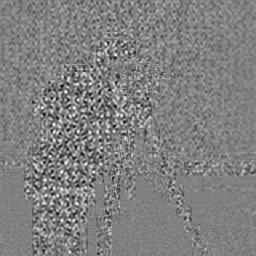}}
\end{minipage}\hfill
\begin{minipage}[c]{.2\linewidth}
\centering{\includegraphics[width = .975\linewidth]{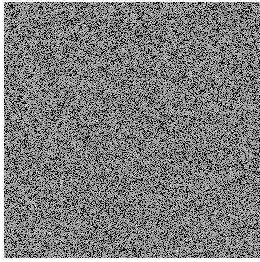}}
\end{minipage}\hfill
\begin{minipage}[c]{.2\linewidth}
\centering{\includegraphics[width = .95\linewidth]{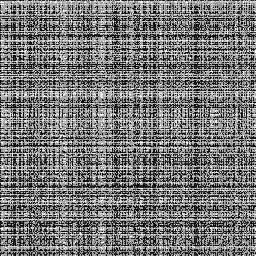}}
\end{minipage}\hfill
\begin{minipage}[c]{.2\linewidth}
\centering{\includegraphics[width = .95\linewidth]{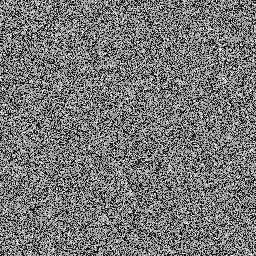}}
\end{minipage}\hfill
\end{minipage}\hfill
\begin{minipage}[c]{1\linewidth}
\begin{minipage}[c]{.2\linewidth}
\centering{\includegraphics[width = .95\linewidth]{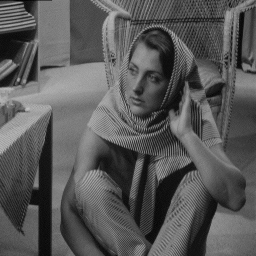}}
\end{minipage}\hfill
\begin{minipage}[c]{.2\linewidth}
\centering{\includegraphics[width = .95\linewidth]{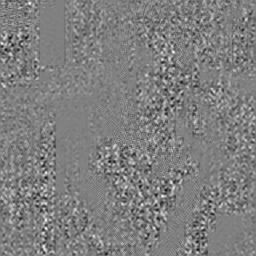}}
\end{minipage}\hfill
\begin{minipage}[c]{.2\linewidth}
\centering{\includegraphics[width = .975\linewidth]{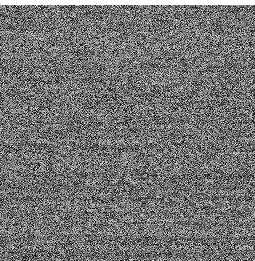}}
\end{minipage}\hfill
\begin{minipage}[c]{.2\linewidth}
\centering{\includegraphics[width = .95\linewidth]{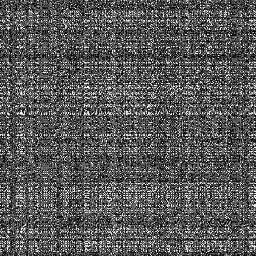}}
\end{minipage}\hfill
\begin{minipage}[c]{.2\linewidth}
\centering{\includegraphics[width = .95\linewidth]{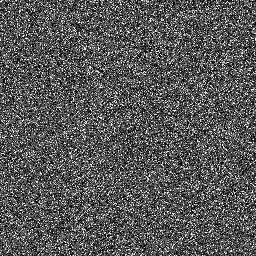}}
\end{minipage}\hfill
\end{minipage}\hfill
\begin{minipage}[c]{1\linewidth}
\begin{minipage}[c]{.2\linewidth}
\centering{\includegraphics[width = .95\linewidth]{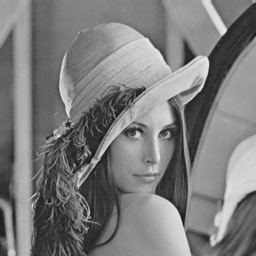}}
\end{minipage}\hfill
\begin{minipage}[c]{.2\linewidth}
\centering{\includegraphics[width = .95\linewidth]{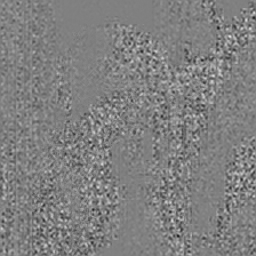}}
\end{minipage}\hfill
\begin{minipage}[c]{.2\linewidth}
\centering{\includegraphics[width = .985\linewidth]{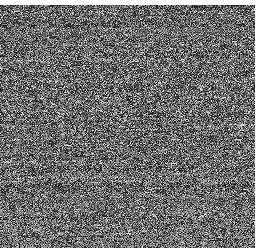}}
\end{minipage}\hfill
\begin{minipage}[c]{.2\linewidth}
\centering{\includegraphics[width = .95\linewidth]{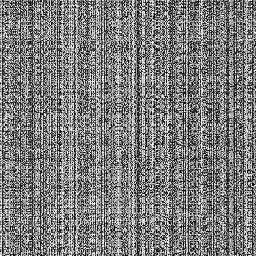}}
\end{minipage}\hfill
\begin{minipage}[c]{.2\linewidth}
\centering{\includegraphics[width = .95\linewidth]{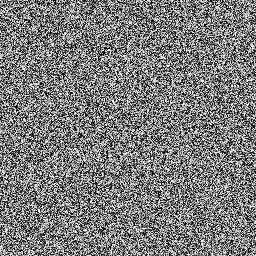}}
\end{minipage}\hfill
\end{minipage}\hfill
\begin{minipage}[c]{1\linewidth}
\begin{minipage}[c]{.2\linewidth}
\centering{\includegraphics[width = .95\linewidth]{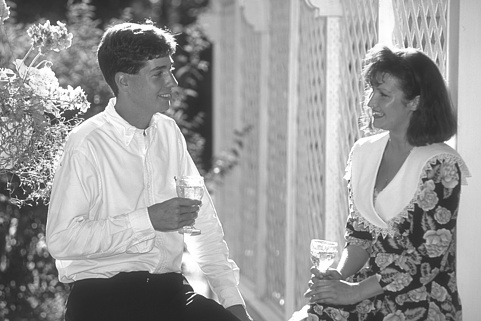}}
\end{minipage}\hfill
\begin{minipage}[c]{.2\linewidth}
\centering{\includegraphics[width = .95\linewidth]{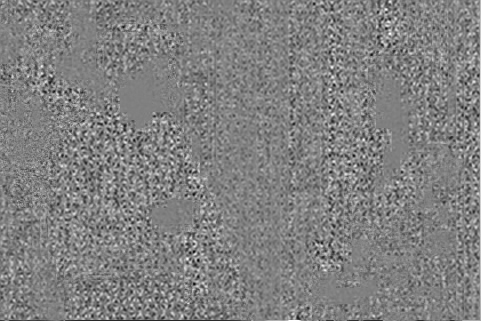}}
\end{minipage}\hfill
\begin{minipage}[c]{.2\linewidth}
\centering{\includegraphics[width = .95\linewidth]{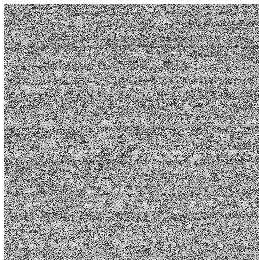}}
\end{minipage}\hfill
\begin{minipage}[c]{.2\linewidth}
\centering{\includegraphics[width = .95\linewidth]{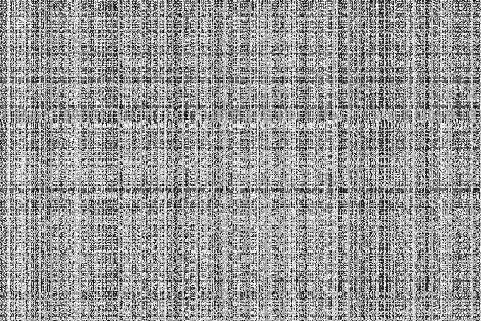}}
\end{minipage}\hfill
\begin{minipage}[c]{.2\linewidth}
\centering{\includegraphics[width = .95\linewidth]{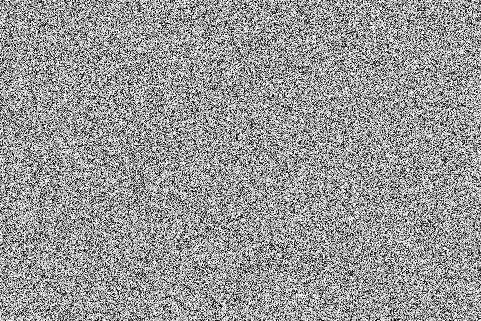}}
\end{minipage}\hfill
\end{minipage}\hfill
\begin{minipage}[c]{1\linewidth}
\begin{minipage}[c]{.2\linewidth}
\centering{\includegraphics[width = .6\linewidth]{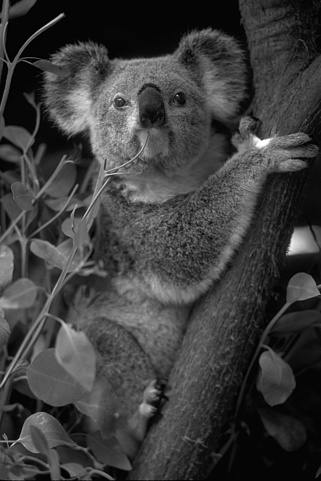}}
\end{minipage}\hfill
\begin{minipage}[c]{.2\linewidth}
\centering{\includegraphics[width = .6\linewidth]{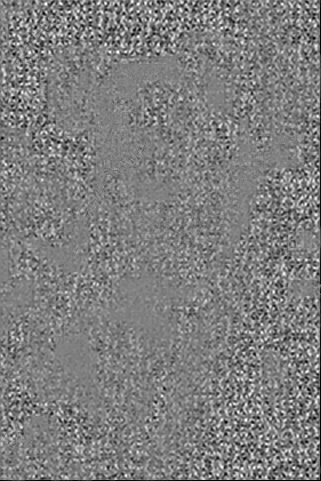}}
\end{minipage}\hfill
\begin{minipage}[c]{.2\linewidth}
\centering{\includegraphics[width = .95\linewidth]{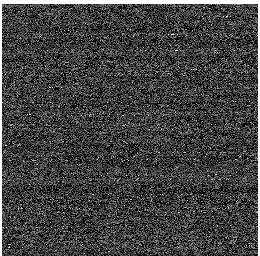}}
\end{minipage}\hfill
\begin{minipage}[c]{.2\linewidth}
\centering{\includegraphics[width = .6\linewidth]{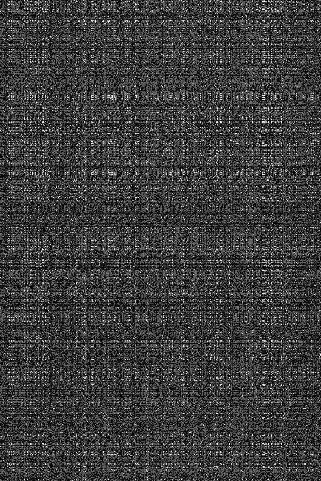}}
\end{minipage}\hfill
\begin{minipage}[c]{.2\linewidth}
\centering{\includegraphics[width = .6\linewidth]{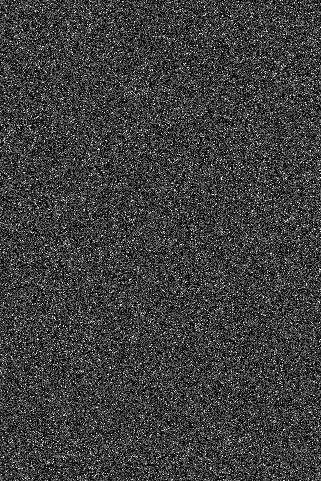}}
\end{minipage}\hfill
\end{minipage}\hfill
\begin{minipage}[c]{1\linewidth}
\begin{minipage}[c]{.2\linewidth}
\centering{\includegraphics[width = .95\linewidth]{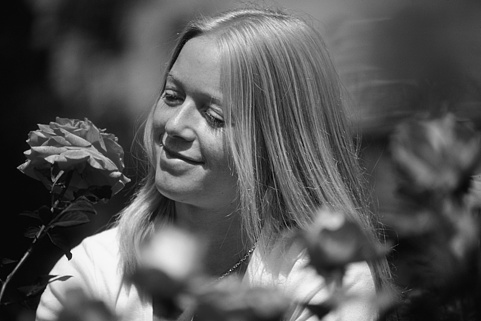} }
\end{minipage}\hfill
\begin{minipage}[c]{.2\linewidth}
\centering{\includegraphics[width = .95\linewidth]{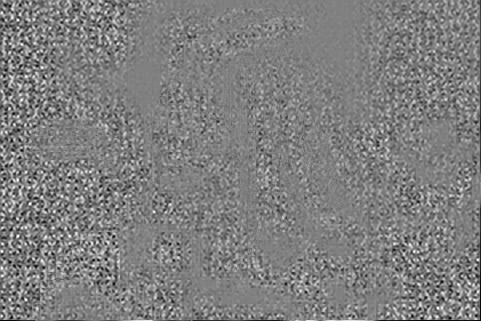} }
\end{minipage}\hfill
\begin{minipage}[c]{.2\linewidth}
\centering{\includegraphics[width = .95\linewidth]{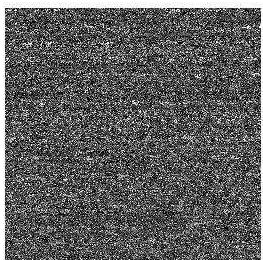}}
\end{minipage}\hfill
\begin{minipage}[c]{.2\linewidth}
\centering{\includegraphics[width = .95\linewidth]{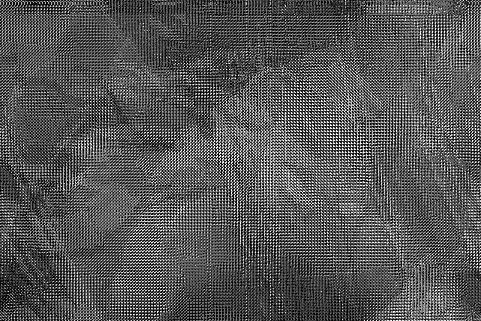}}
\end{minipage}\hfill
\begin{minipage}[c]{.2\linewidth}
\centering{\includegraphics[width = .95\linewidth]{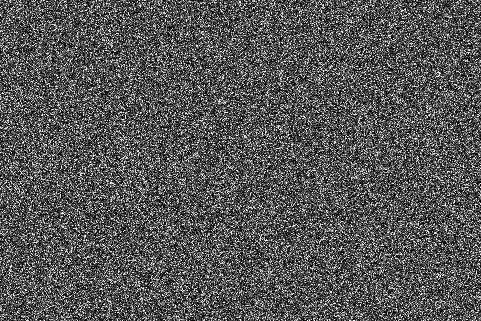}}
\end{minipage}\hfill
\end{minipage}\hfill
\caption{Image scrambling results with visual comparisons; 1st column: test images (from 1st row to the last are \textit{Cameraman}, \textit{Barbara}, \textit{Lenna}, and $\#157055, \#69015$ and $\#239096$); 2nd column: simulation results of Van De Ville \etal's method \cite{VanScrambler}; 3rd column: simulation results of Abu Dalhoum \etal's method \cite{Abu}; 4th column: simulation results of Ye's method \cite{ye2010image}; and 5th column: simulation results of ours.}
\label{fig:simulation}
\end{figure}

Fig. \ref{fig:simulation} shows the simulation results of the proposed image scrambling method using the two dimensional Sudoku associated bijections with peer algorithms \cite{ye2010image,Abu}. As can bee seen from these results, the proposed method outperforms the two recent algorithms, in the sense that its scrambled image is random-like and pixels more evenly scrambled, while results of Van De Ville \etal's method \cite{VanScrambler} contain distinguishable patterns of foreground from background; results of Abu Dalhoum \etal's method \cite{Abu} contain line-like patterns and results of Ye's method \cite{ye2010image} are of mesh-like patterns. Furthermore, it is noticeable that Abu Dalhoum \etal's method might require extra space to deal with the edge effect in cellular automata (see results on the 2nd column), while the chaos map used Ye's method might be not that dynamic and generates weak results (see Ye's result on test image \#239096).

\begin{figure}[!h]
\centering
\scriptsize
\begin{minipage}[c]{.8\linewidth}
\begin{minipage}[c]{.25\linewidth}
\centering{\includegraphics[width = .95\linewidth]{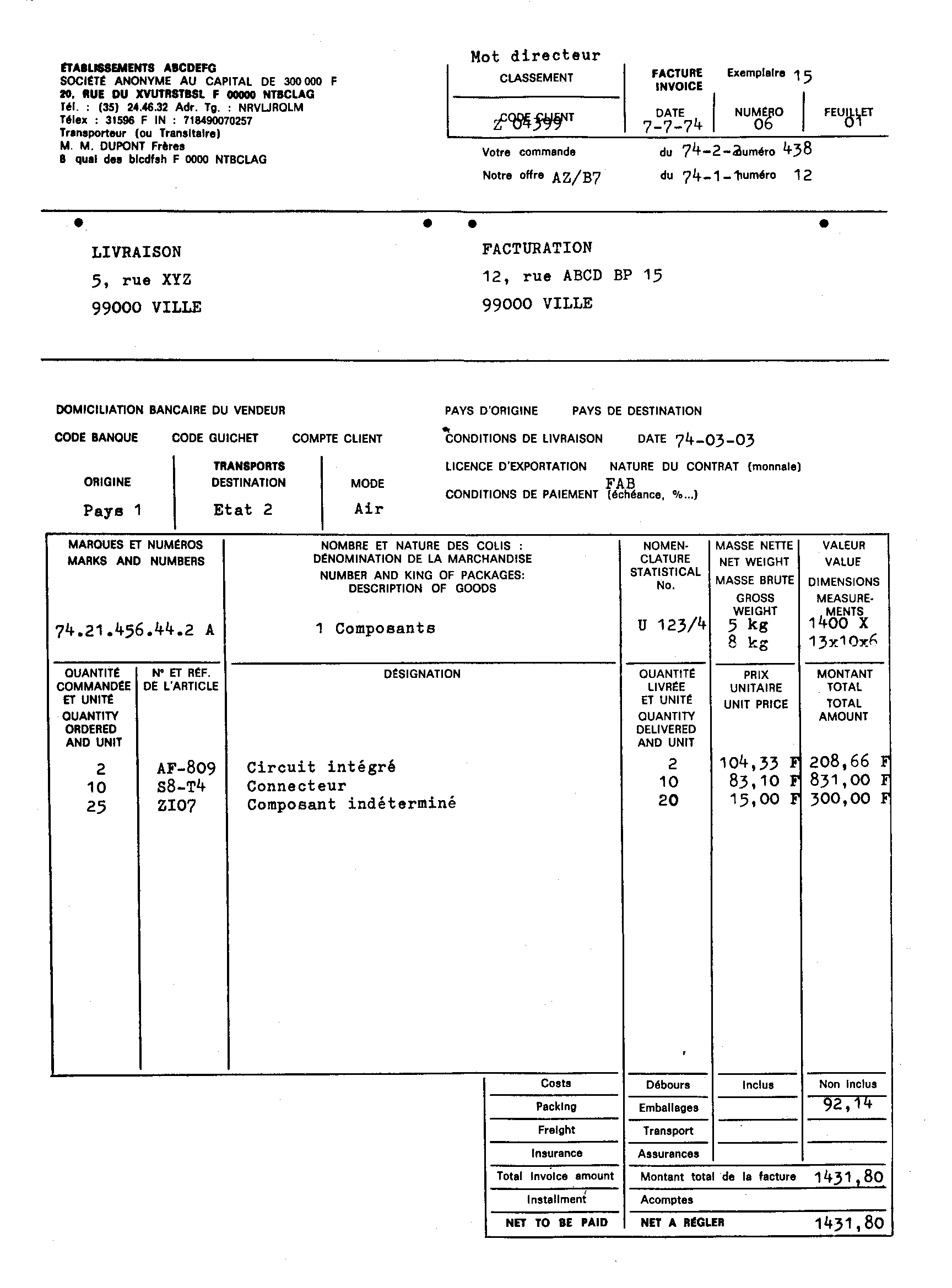}}
\centerline{(a)}
\end{minipage}\hfill
\begin{minipage}[c]{.25\linewidth}
\centering{\includegraphics[width = .975\linewidth]{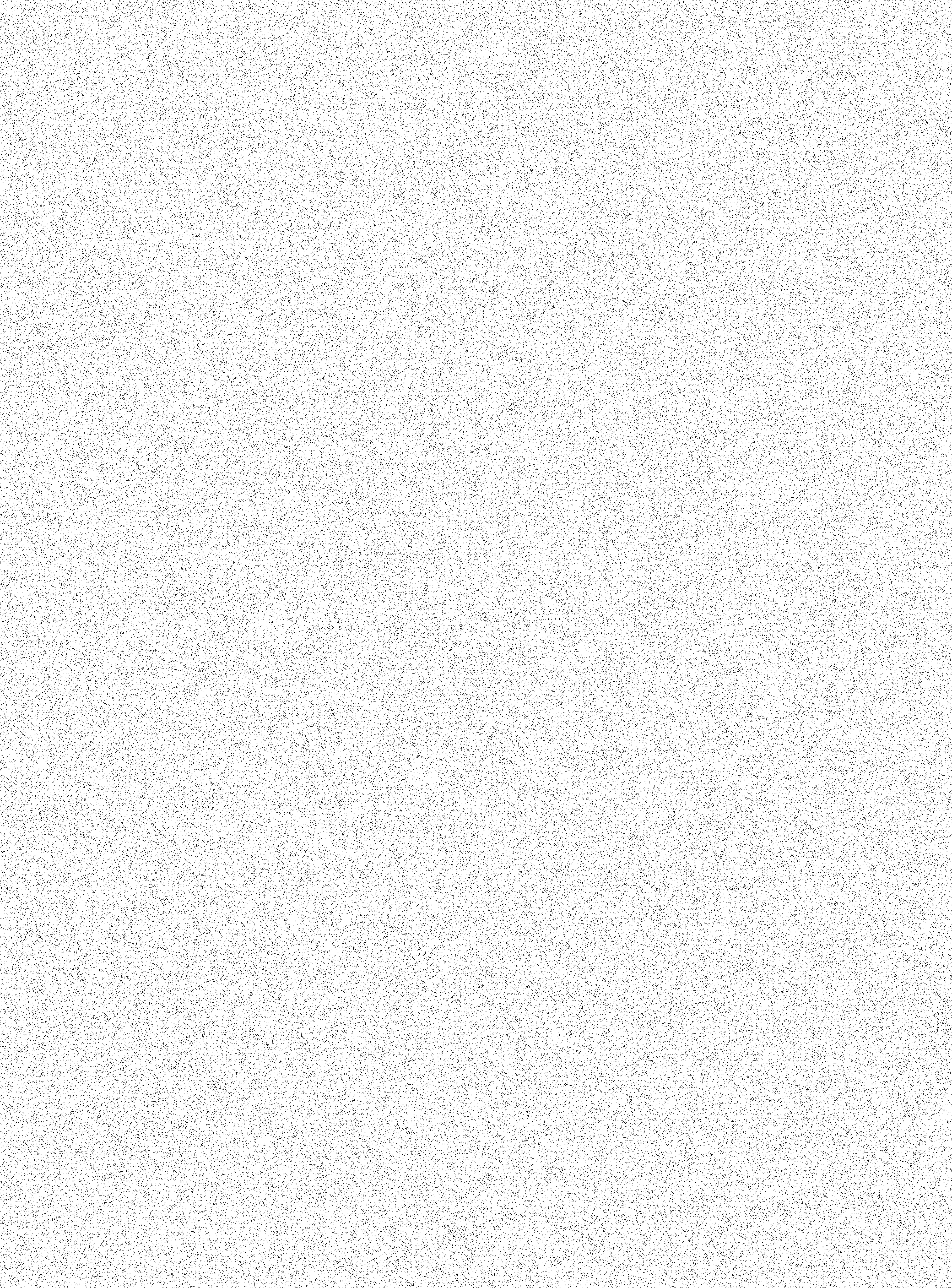}}
\centerline{(b)}
\end{minipage}\hfill
\begin{minipage}[c]{.25\linewidth}
\centering{\includegraphics[width = .95\linewidth]{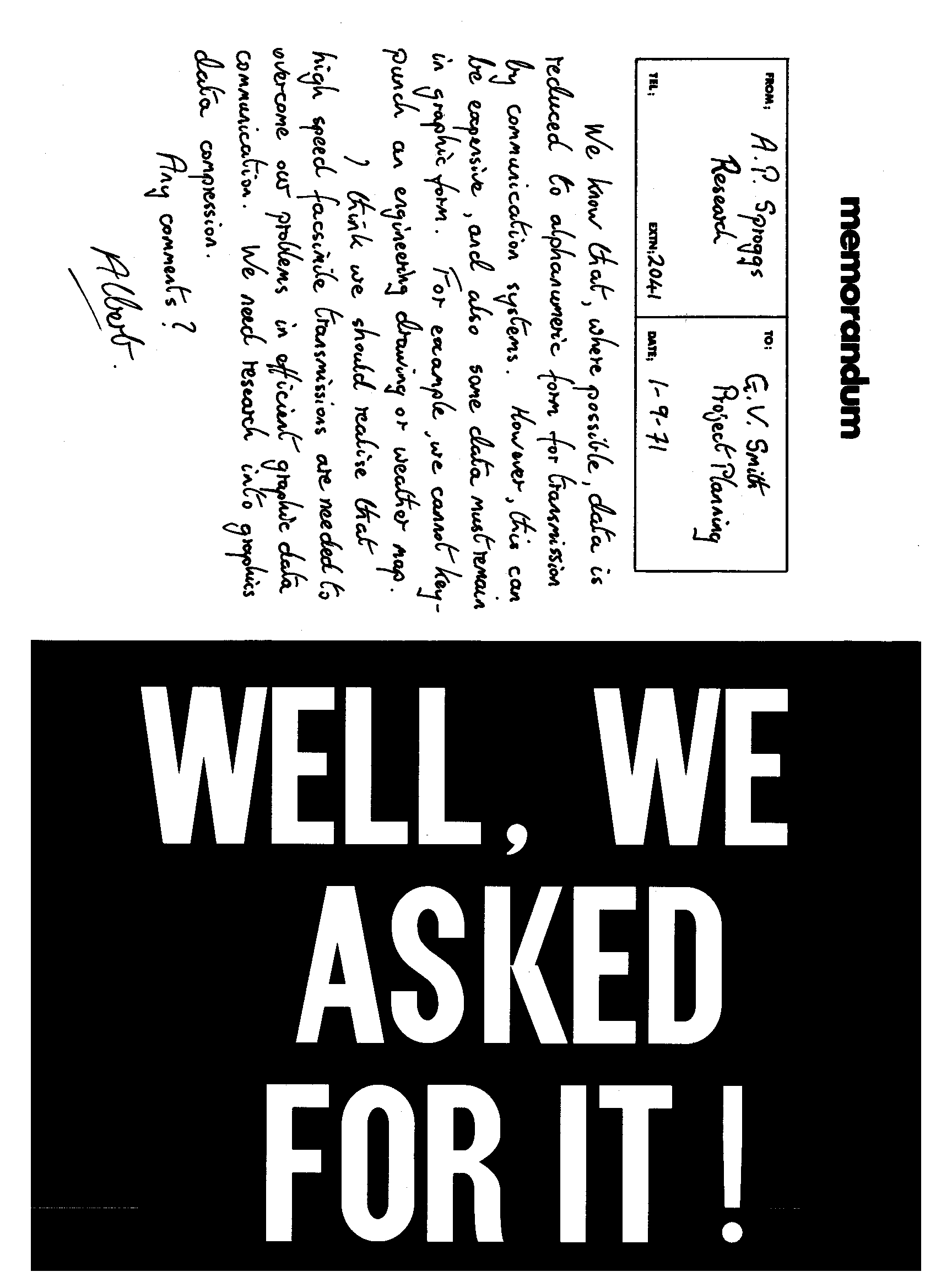}}
\centerline{(c)}
\end{minipage}\hfill
\begin{minipage}[c]{.25\linewidth}
\centering{\includegraphics[width = .95\linewidth]{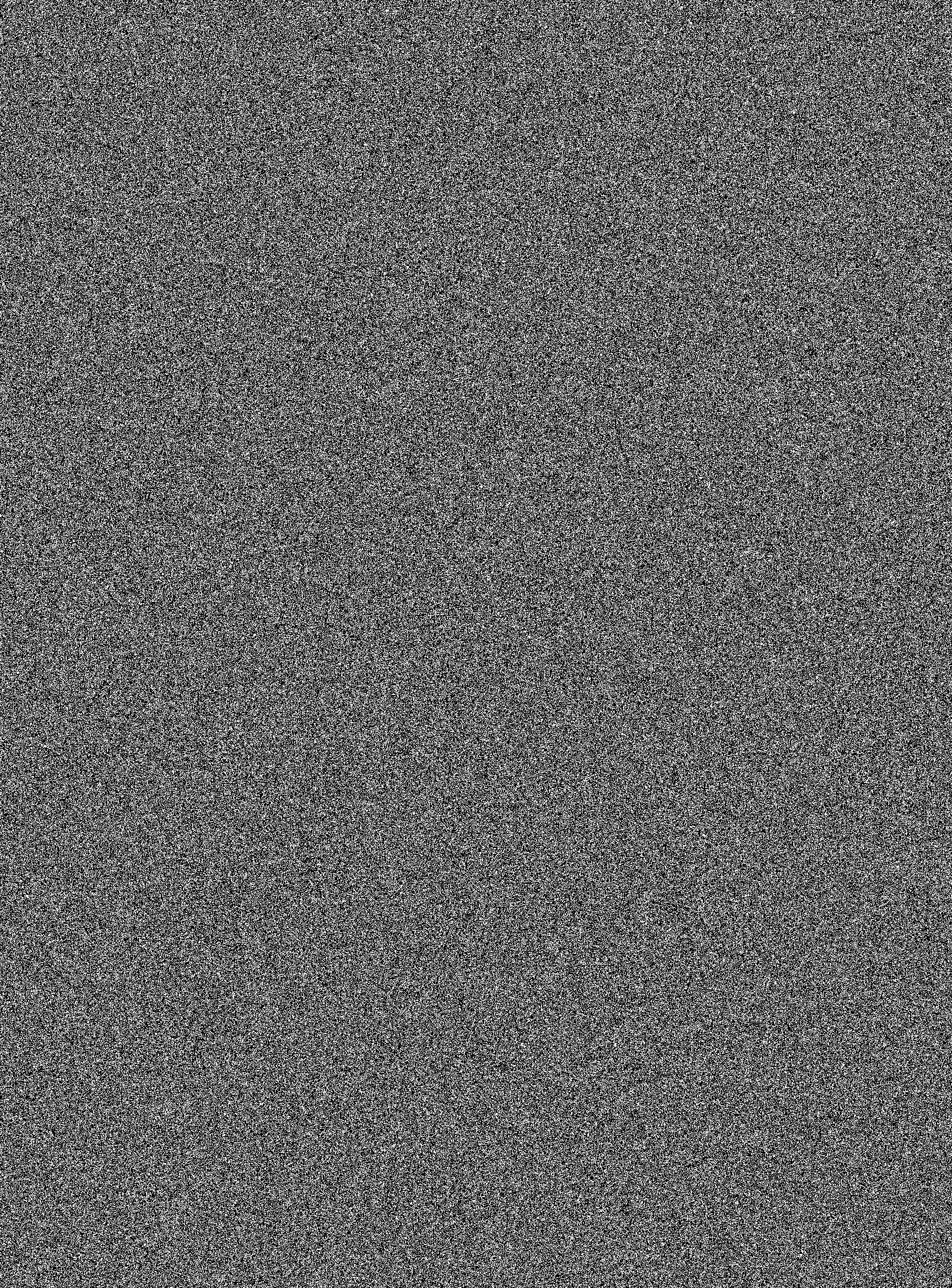}}
\centerline{(d)}
\end{minipage}\hfill
\end{minipage}\hfill
\begin{minipage}[c]{.8\linewidth}
\begin{minipage}[c]{.25\linewidth}
\centering{\includegraphics[width = .95\linewidth]{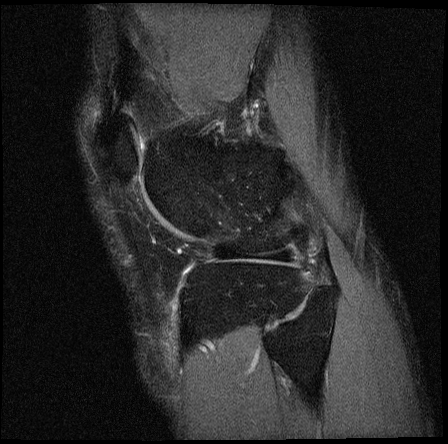}}
\centerline{(e)}
\end{minipage}\hfill
\begin{minipage}[c]{.25\linewidth}
\centering{\includegraphics[width = .975\linewidth]{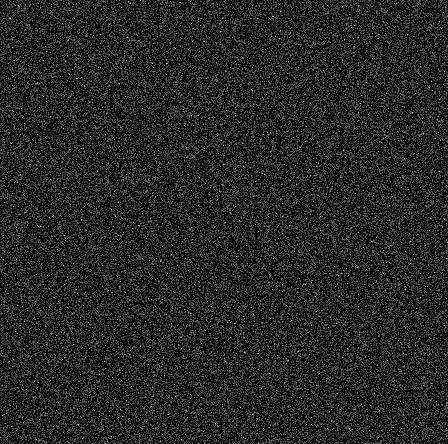}}
\centerline{(f)}
\end{minipage}\hfill
\begin{minipage}[c]{.25\linewidth}
\centering{\includegraphics[width = .95\linewidth]{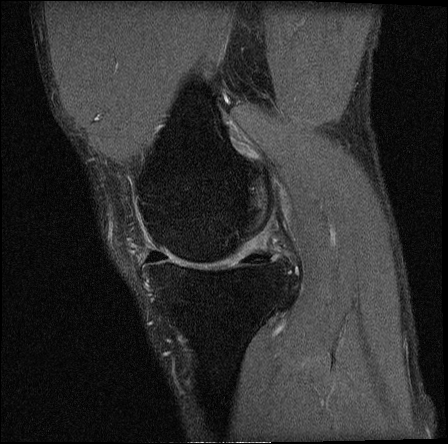}}
\centerline{(g)}
\end{minipage}\hfill
\begin{minipage}[c]{.25\linewidth}
\centering{\includegraphics[width = .95\linewidth]{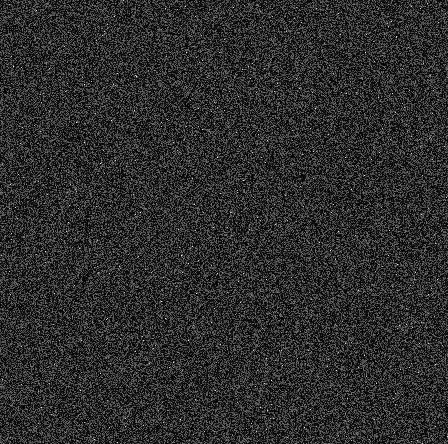}}
\centerline{(h)}
\end{minipage}\hfill
\end{minipage}\hfill
\begin{minipage}[c]{.8\linewidth}
\begin{minipage}[c]{.25\linewidth}
\centering{\includegraphics[width = .95\linewidth]{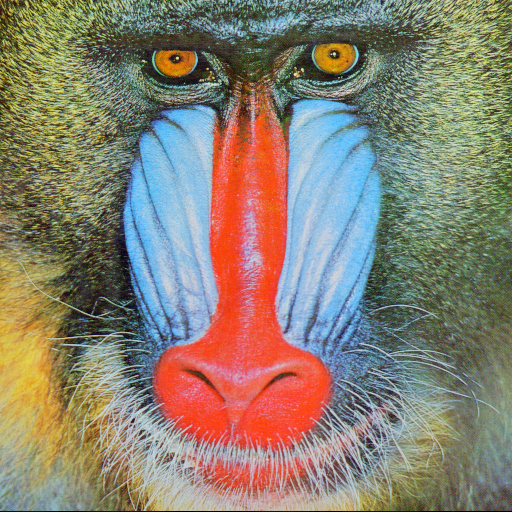}}
\centerline{(i)}
\end{minipage}\hfill
\begin{minipage}[c]{.25\linewidth}
\centering{\includegraphics[width = .975\linewidth]{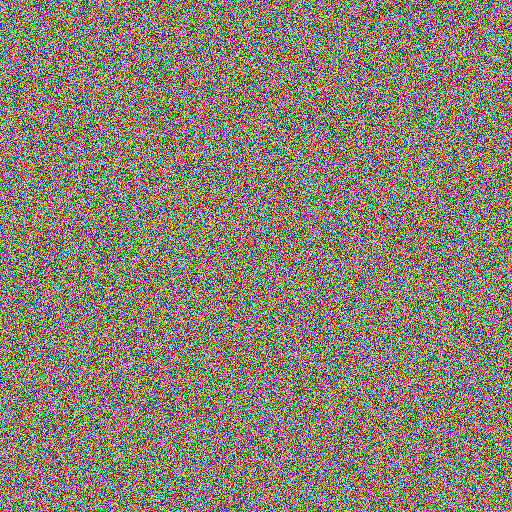}}
\centerline{(j)}
\end{minipage}\hfill
\begin{minipage}[c]{.25\linewidth}
\centering{\includegraphics[width = .95\linewidth]{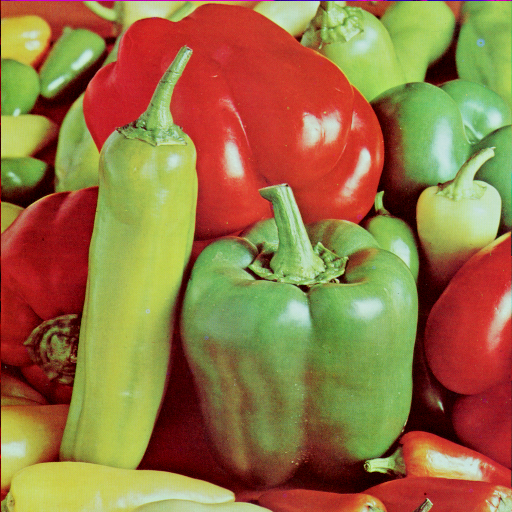}}
\centerline{(k)}
\end{minipage}\hfill
\begin{minipage}[c]{.25\linewidth}
\centering{\includegraphics[width = .95\linewidth]{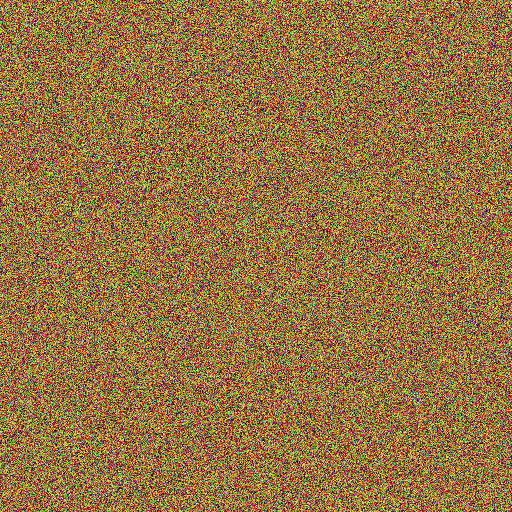}}
\centerline{(l)}
\end{minipage}\hfill
\end{minipage}\hfill
\caption{Image scrambling results on various image types using Sudoku associated image scrambler; (a) and (b) \textit{CCITT-3} before and after scrambling; (c) and (d) \textit{CCITT-7} before and after scrambling; (e) and (f) \textit{knee MRI sample A} before and after scrambling; (g) and (h) \textit{knee MRI sample B} before and after scrambling; (i) and (j) \textit{4.2.03} before and after scrambling, and (k) and (l) \textit{4.2.07} before and after scrambling.}
\label{fig:simulation2}
\end{figure}

Besides the 8-bit grayscale images listed in Fig. \ref{fig:simulation}, we also test the proposed method on various image types including binary images \textit{CCITT-3} and \textit{CCITT-7} from the CCITT fax image compression dataset \footnotemark[4], 16-bit grayscale knee MRI images, and color images \textit{4.2.03} and \textit{4.2.07} from the USC-SIPI miscellaneous image dataset \footnotemark[5]. Scrambling results are given in Fig. \ref{fig:simulation2}.

In regard of the detail information about these test images including processing time and speed, we list these relevant information in Table \ref{tab:time}. Roughly speaking, the speed of the proposed scrambler is about $150KB/s$, or $0.15MB/s$ equivalently under MATLAB environment. It is worthwhile to note that this speed can be largely enhanced by using parallel computations, because the scrambling process of each bit-plane is independent of each other and image blocks that not overlapped are also independent of each other, both of which implying these works can be done in parallel to save time. Meanwhile, it is well-known that MATLAB is very slow for \textit{for-loop} execution, and thus implementing the proposed scrambler in other languages might further enhance the processing speed. Therefore, the proposed scrambler with a proper implement could meet the demand for real time processings.

\begin{table}[h]
\caption{Test Images Information with Execution Speed}\label{tab:time}
\centering
\scriptsize
\begin{tabular}{r|rrr|rr}
  \hline\hline
& \multicolumn{3}{c|}{\textbf{Test Image Information}}&\multicolumn{2}{c}{\textbf{Execution Information}}\\
	\textbf{Image File} & \textbf{Width} &\textbf{Height} &\textbf{Bit Depth} &	 \textbf{Time(s)} & \textbf{Speed(KB/s)}\\\hline
\textbf{\textit{Cameraman}}&	256	&256&	8&	2.8280	&181.0467\\
\textbf{\textit{Barbara}}	&256&	256&	8	&2.9266	&174.9470\\
\textbf{\textit{Lenna}}	&256	&256	&8	&3.0336	&168.7764\\
\textbf{\textit{\#157055}}&	321&	481	&8	&5.0176	&240.4053\\
\textbf{\textit{\#69015}}	&481&	321	&8	&4.6216&	261.0044\\
\textbf{\textit{\#239096}}	&321&	481&	8	&4.6537	&259.2040\\\hline
\textbf{\textit{CCITT-3}}	&2339&	1728	&1	&31.4804	&125.3816\\
\textbf{\textit{CCITT-7}}&	2339	&1728	&1	&32.8107	&120.2980\\
\textbf{\textit{Knee MRI Sample A}}&	448	&448	&16	&21.1253	&148.4476\\
\textbf{\textit{Knee MRI Sample B}}&	448	&448	&16	&21.1838	&148.0377\\
\textbf{\textit{4.2.03}}	&512	&512	&24	&31.1277	&197.3805\\
\textbf{\textit{4.2.07}}	&512&	512	&24	&32.4009	&189.6244\\
  \hline\hline
\end{tabular}
\end{table}

\subsection{Gray Difference and Gray degree of Scrambling}
Gray difference and gray degree of scrambling (GDD) \cite{ye2008novel} are two measures used for quality evaluation of scrambled images. The gray difference for a $W\times H$ image $X$ is defined as follows,
\begin{equation}\label{eq:GD}
    GD^X_{i,j} = \dfrac{1}{4}\sum\limits_{p,q \in \{-1,+1\}}{\left(X_{i,j}-X_{i+p,j+q}\right)^2}
\end{equation}
Further, the mean gray difference of image $X$ is computed by averaging all pixels except those on image edges as shown in Eq. \eqref{eq:EGD}.
\begin{equation}\label{eq:EGD}
    EGD^X_{i,j} = \dfrac{\sum\limits_{i=2}^{W-1}\sum\limits_{j=2}^{H-1}{GD^X_{i,j}}}{(W-2)(H-2)}
\end{equation}
Finally, the gray degree of scrambling is defines as the ratio of the difference of the mean gray differences before and after scrambling image $X$ to their sum, namely
\begin{equation}\label{eq:EGD}
    GGD^{X,X'}_{i,j} = \dfrac{|EGD^X_{i,j}-EGD^{X'}_{i,j}|}{EGD^X_{i,j}+EGD^{X'}_{i,j}}
\end{equation}

Table \ref{tab:GDD} compares the gray degree of scrambling of the proposed method with peer algorithms \cite{ye2007image,VanScrambler,ye2008novel,Abu,ye2010image}\footnotemark[4]. As can be seen, the proposed method outperforms the listed peer algorithm by achieving a higher gray degree of scrambling.
\begin{table}[h]
\caption{Gray Scrambling Degree of Scrambled Images}\label{tab:GDD}
\scriptsize
\begin{tabular}{r|cccccc}
  \hline\hline
  &\multicolumn{6}{c}{\textbf{Test Images}}\\
  \textbf{Method}& \textbf{\textit{Cameraman}}& \textbf{\textit{Barbara}} & \textbf{\textit{Lenna}} & \textbf{\#157055}& \textbf{\#69015} & \textbf{\#239096} \\\hline
  \textbf{Van De Ville \etal's \cite{VanScrambler}, 2004} & 0.1315 & 0.1930 & 0.1353 & 0.2752 & 0.5397 & 0.4502 \\
  \textbf{Ye \etal's \cite{ye2007image}, 2007} & 0.8832 & $N\backslash A$ & 0.9010 & 0.8780&0.8646&0.8976\\
  \textbf{Ye \etal's \cite{ye2008novel}, 2008} & 0.8926  & 0.8740  &  0.9311  &  0.8731 & 0.8789 & 0.9134\\
  \textbf{Abu Dalhoum \etal's \cite{Abu}, 2012}& 0.8971 &  0.8749 &  0.9320  & 0.8821 & 0.8827 & 0.9388\\
  \textbf{Ye's \cite{ye2010image}, 2010}  & 0.9118     & 0.9176     & 0.9618    &  0.9218   &   0.9406   &   0.9670 \\
  \textbf{Ours} & 0.9165    &  0.9200  &    0.9663   &   0.9359 &    0.9439  &   0.9672\\
  \hline\hline
\end{tabular}
\end{table}

\footnotetext[4]{Results of {Ye \etal's \cite{ye2007image},2007}, {Ye \etal's \cite{ye2008novel}, 2008} and Abu Dalhoum \etal's \cite{Abu} are directly taken from listed results in \cite{Abu}.}

\subsection{Adjacent Pixel Autocorrelation}
\begin{figure}
\scriptsize
\centering
\begin{minipage}[c]{.5\linewidth}
    \includegraphics[width = .95\linewidth]{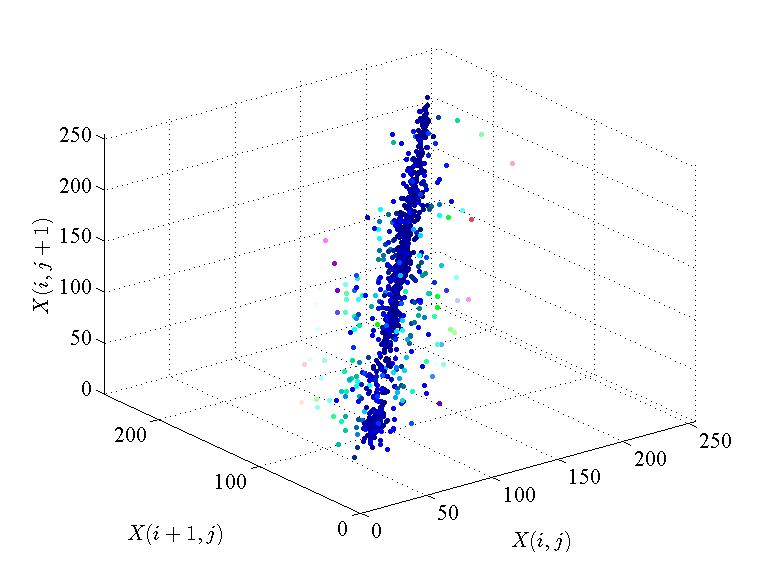}
    \centerline{(a)}
\end{minipage}\hfill
\begin{minipage}[c]{.5\linewidth}
    \includegraphics[width = .95\linewidth]{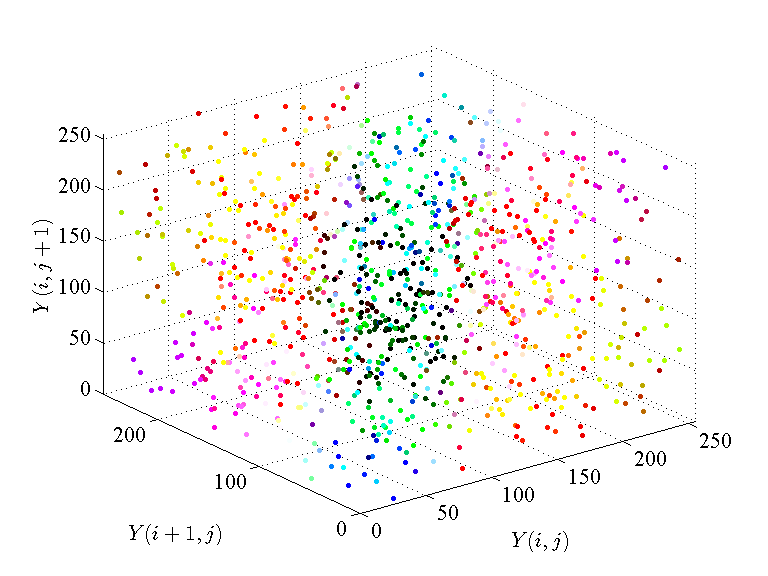}
    \centerline{(b)}
\end{minipage}\hfill
  \caption{Adjacent pixel correlations before and after Sudoku associated image scrambling for image \textit{Lenna}; (a) before scrambling; and (b) after scrambling. Point colors reflect the distance between a point and the line $x=y=z$. The closer a point to this line, the bluer the color is.}\label{fig:apcc}
\end{figure}

A digital image is commonly of high information redundancy in the sense that adjacent pixels are strongly correlated. In contrast, a well scrambled image should break such correlations between neighbor pixels, so that the scrambled image is unrecognizable. For example, one can simply plot a pixel correlation figure as shown in Fig. \ref{fig:apcc}, where each subfigure illustrating the adjacent pixel correlations before and after scrambling image \textit{Lenna} by plotting the intensity values of 1024 random select pixels as $x$ coordinates, intensity values of their horizontal neighbors as $y$ coordinates, and intensity values of their vertical neighbors as $z$ coordinates. It is obvious that before scrambling adjacent pixels in image \textit{Lenna} are close correlated and thus the plots concentrates on the diagonal direction, namely the line $x=y=z$. In contrast, after scrambling the plot of the same set of adjacent pixels speared almost everywhere.

Adjacent pixel autocorrelation coefficient (APCC) is a common measure used in signal process and image encryption \cite{wuSPIE,wuIEEE,Ruisong2011,wu2011wheel,Zhang2011,zhu2010chaos}. Mathematically, we can define this autocorrelation coefficient as follows:
\begin{equation}
    \label{EqCorrd}
    {\rho} = {\textrm{E}}[(X_t-\mu)(X_{t+1}-\mu)]/\sigma^2
\end{equation}
where $X_t$ denotes the $t$th element in signal $X$, ${\textrm{E}}(.)$ denotes the mathematica expectation as shown in Eq. \eqref{EqExpectation}, $\mu$ is the expectation namely the mean value defined in Eq. \eqref{EqMean}, and $\sigma$ denotes the standard deviation defined in Eq. \eqref{EqStd}.
\begin{equation}
    \label{EqExpectation}
    {\textrm{E}}[Y] = \sum_{i=1}^{N}Y_i/N
\end{equation}
\begin{equation}
    \label{EqMean}
    \mu = {\textrm{E}}[X]
\end{equation}
\begin{equation}
    \label{EqStd}
    \sigma = \sqrt{{\textrm{E}}[(X-\mu)^2]}
\end{equation}
Consequently, less correlated adjacent pixels within image $X$ are, the closer ${\rho}$ approaches zero. In contrast, if adjacent pixels within image $X$ are completely dependent, then $|{\rho}| = 1$.

Commonly the adjacent pixels in image can be defined along the horizontal direction and the vertical direction, respectively. Equivalently, we extract scrambled image pixels along rows and along columns respectively and compare the autocorrelation coefficients for these two pixel sequences. Results are given in Tables \ref{tab:APCh} and \ref{tab:APCv}. It is not difficult to see that the proposed method outperforms the three peer algorithms again.
\begin{table}[h]
\caption{Horizontal Adjacent Pixel Correlation Coefficients of Scrambled Images}\label{tab:APCh}
\scriptsize
\begin{tabular}{r|cccccc}
  \hline\hline
  \textbf{Horizontal APCC} &\multicolumn{6}{c}{\textbf{Test Images}}\\
  \textbf{Method}& \textbf{\textit{Cameraman}}& \textbf{\textit{Barbara}} & \textbf{\textit{Lenna}} & \textbf{\#157055}& \textbf{\#69015} & \textbf{\#239096} \\
  \textbf{Before Scrambling} &    0.9565   & 0.9238 &   0.9709&    0.9522  &  0.9613  &  0.9796\\\hline
  \textbf{Van De Ville \etal's \cite{VanScrambler}, 2004} & 0.6967 & 0.6974 & 0.7156 & 0.6963 & 0.7099 & 0.7005 \\
    \textbf{Abu Dalhoum \etal's \cite{Abu}, 2012}& 0.0184 &  -0.0629 &  -0.0204  & -0.0254 & 0.1564 & 0.1771\\
  \textbf{Ye's \cite{ye2010image}, 2010}  &   0.0827   & 0.0398  &  0.2130  &  0.2270   & 0.0359 &  -0.0372\\
  \textbf{Ours} &  -0.0020   & -0.0015   & -0.0021  &  0.0024  &  -0.0025 &   -0.0017\\
  \hline\hline
\end{tabular}
\end{table}

\begin{table}[h]
\caption{Vertical Adjacent Pixel Correlation Coefficients of Scrambled Images}\label{tab:APCv}
\scriptsize
\begin{tabular}{r|cccccc}
  \hline\hline
  \textbf{Vertical APCC} &\multicolumn{6}{c}{\textbf{Test Images}}\\
  \textbf{Method}& \textbf{\textit{Cameraman}}& \textbf{\textit{Barbara}} & \textbf{\textit{Lenna}} & \textbf{\#157055}& \textbf{\#69015} & \textbf{\#239096} \\
  \textbf{Before Scrambling} &     0.9333  &  0.8797    &0.9400   & 0.9475   & 0.9570 &   0.9739\\\hline
  \textbf{Van De Ville \etal's \cite{VanScrambler}, 2004} & 0.6613 & 0.6699 & 0.6741 & 0.6690 & 0.6785 & 0.6865 \\
    \textbf{Abu Dalhoum \etal's \cite{Abu}, 2012}& 0.0543& 0.0172    &0.0476    &0.0376&    0.2162   & 0.2314\\
  \textbf{Ye's \cite{ye2010image}, 2010} &  0.2808  &   0.0598  &  -0.0955  &   0.1015 &    0.0860  &  -0.0329\\
  \textbf{Ours} &  -0.0033  & -0.0031 &   0.0016&   -0.0004   & 0.0012   & 0.0040\\
  \hline\hline
\end{tabular}
\end{table}
It is also well-known that whether an observed autocorrelation coefficient ${\rho}$ is significantly different from zero can be tested by \textit{Student's t-distribution} \cite{donner1980inferences,macklin1982investigation,lowry2005concepts}. Specifically in \cite{lowry2005concepts}, it states that ``if the true correlation between $X$ and $Y$ within the general population is zero, and if the size of sample $T$, on which an observed value $\rho$ is based is equal to or greater than 6, then the quantity $t$ defined by Eq. \eqref{eq:t}
\begin{equation}\label{eq:t}
    t = {\rho}\sqrt{\dfrac{T-2}{1-{\rho}^2}}
\end{equation}
is distributed approximately as a \textit{Student's t-distribution} with $T-2$ degrees of freedom
'', where \textit{Student's t-distribution} has the probability density function given by Eq. \eqref{eq:pdft}, where $v$ is the number of degrees of freedom and $\Gamma(.)$ is the Gamma function.
\begin{equation}\label{eq:pdft}
    g(t) = \dfrac{\Gamma\left({v+1\over 2}\right)}{\sqrt{v\pi}\Gamma\left({v\over 2}\right)}\left(1+{t^2\over v}\right)^{-{v+1\over 2}}
\end{equation}
Therefore, given an autocorrelation coefficient of an $W\times H$ image $X$, it is not difficult to verify whether its adjacent pixels are correlated or not by taking the test statistic $t$ for two-sided hypothesis tests. Specifically, one can calculate the P-value of a sample correlation coefficient $\rho$ by finding its P-Value under the null hypothesis that the derived statistic $t$ from $\rho$ follows a \textit{Student's t-distribution}, where the P-value of a given test statistic $t$ is computed as follows
\begin{equation}\label{eq:pvalue}
    \textrm{P-Value}(t) = 2\int_{-\infty}^{-|t|} g(\tau) d\tau
\end{equation}
Statistically speaking, a P-Value is a measure of how much evidence we have under the null hypothesis. In other words, the smaller P-Value the more evidence we have against the hypothesis. The range of a P-Value is $[0,1]$.

In our situation, the sample size $T$ is the number of pixels within image $X$, which is $T = WH$. Consequently, test statistic $t$ can be derived by the autocorrelation coefficient $\rho$ and $T$ using Eq. \eqref{eq:t}. As a result, corresponding P-value results can be calculated for the autocorrelation coefficients of the scrambled image using the proposed method. These results are shown in Table \ref{tab:pvalue}. It is clear that these P-Values are much greater than $5\%$, which is a common critical value used in statistics implying to reject the null hypothesis if a P-Value is less than $5\%$, while accept the null hypothesis otherwise. Therefore, these results indicate that the correlation coefficient between adjacent pixels in the scrambled image using the proposed method are indeed zeros. In other words, the proposed method successfully breaks the strong correlations between adjacent pixels after scrambling.

\begin{table}[h]
\caption{P-Values of Adjacent Pixel Correlation Coefficients}\label{tab:pvalue}
\scriptsize
\begin{tabular}{r|cccccc}
  \hline\hline
  \textbf{} &\multicolumn{6}{c}{\textbf{Test Images}}\\
  \textbf{Statistics}& \textbf{\textit{Cameraman}}& \textbf{\textit{Barbara}} & \textbf{\textit{Lenna}} & \textbf{\#157055}& \textbf{\#69015} & \textbf{\#239096} \\
  \textbf{Size} &     $256\times 256$ &  $256\times 256$   &$256\times 256$ & $481\times 321$  & $321\times 481$ &   $481\times 321$ \\
  \textbf{Degree of Freedom $v$} & 65534 & 65534 & 65534 & 154399& 154399& 154399\\\hline
  \textbf{APCC $\rho_{horizontal}$} &  -0.0020   & -0.0015   & -0.0021  &  0.0024  &  -0.0025 &   -0.0017\\
  \textbf{Statistic $t_{horizontal}$} & -0.5120 & -0.3840 & -0.5376 & 0.9431 & -0.9823& -0.6680 \\
  \textbf{P-Value $ P_{horizontal}$} & 60.84\% & 70.10\% & 58.08\% & 34.56\% & 32.60\% & 50.42\% \\\hline
  \textbf{APCC $\rho_{vertical}$} &  -0.0033  & -0.0031 &   0.0016&   -0.0004   & 0.0012   & 0.0040\\
  \textbf{Statistic $t_{vertical}$} & -0.8448 & -0.7936 & 0.4096 & 0.1572 & 0.4715& -1.5718 \\
  \textbf{P-Value$  P_{vertical}$} & 37.62\% & 42.74\% & 68.22\% & 87.50\% & 63.62\% & 11.60\% \\
  \hline\hline
\end{tabular}
\end{table}
\subsection{Key Space and Key Sensitivity}
A good image scrambler should have a sufficiently large key space to resist \textit{brute-force} attacks. \textit{Sudoku Associated Image Scrambler} is designed to be of 192-bit length, which is considered sufficiently large to be immune to this type of attacks with respect to the current computer capacities. on the other hand, this key space can be easily extended because the number of possible Sudoku matrices are extremely huge. A lower bound of the number of $256\times 256$ Sudoku matrices is $256! \approx 2^{1684}$. Considering that majority of digital image files are larger than $256\times 256$, the number of Sudoku matrices at larger sizes are even huge.

\begin{figure}[h]
\centering
\scriptsize
\begin{minipage}[c]{.8\linewidth}
\begin{minipage}[c]{.33\linewidth}
\centering{\includegraphics[width = .9\linewidth]{4_2_03}}
\centerline{(a)}
\end{minipage}\hfill
\begin{minipage}[c]{.33\linewidth}
\centering{\includegraphics[width = .9\linewidth]{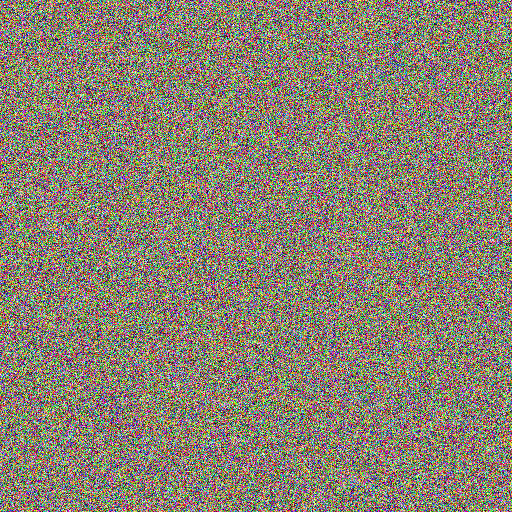}}
\centerline{(b)}
\end{minipage}\hfill
\begin{minipage}[c]{.33\linewidth}
\centering{\includegraphics[width = .9\linewidth]{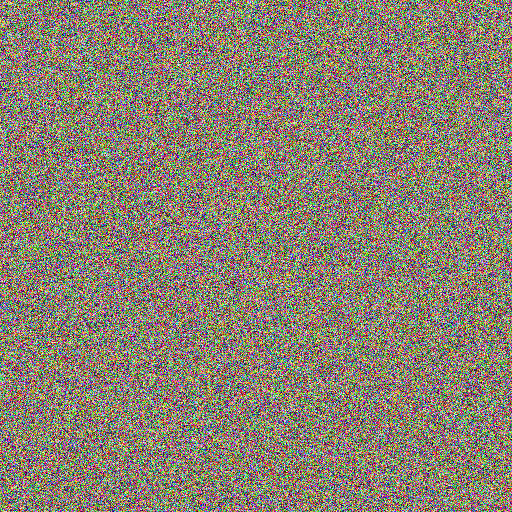}}
\centerline{(c)}
\end{minipage}\hfill
\begin{minipage}[c]{.33\linewidth}
\centering{\includegraphics[width = .9\linewidth]{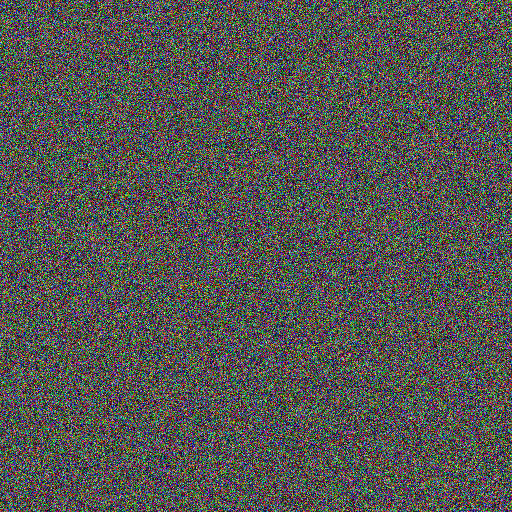}}
\centerline{(d)}
\end{minipage}\hfill
\begin{minipage}[c]{.33\linewidth}
\centering{\includegraphics[width = .9\linewidth]{4_2_03}}
\centerline{(e)}
\end{minipage}\hfill
\begin{minipage}[c]{.33\linewidth}
\centering{\includegraphics[width = .9\linewidth]{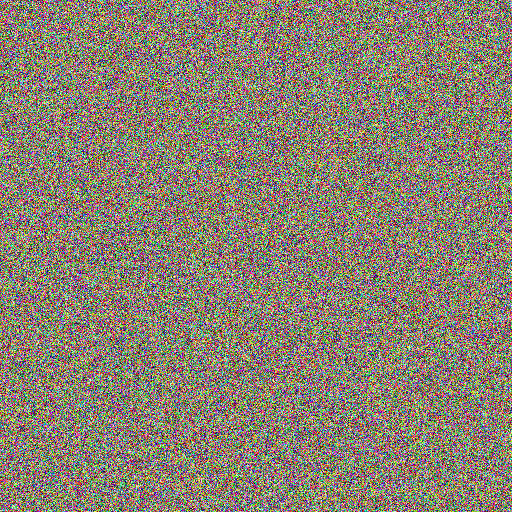}}
\centerline{(f)}
\end{minipage}\hfill
\end{minipage}\hfill
\caption{Key sensitivities of \textit{Sudoku Associated Image Scrambler}; (a) image \textit{4.02.03}; (b) scrambled image $Y_a$ using $K_a$; (c) scrambled image $Y_b$ using $K_b$; (d) difference image of $|Y_a-Y_b|$; (e) descrambled image of $Y_a$ using $K_a$; and (f) descrambled image of $Y_a$ using $K_b$.}
\label{fig:sensitivity}
\end{figure}

In addition, a good scrambler should have high sensitivity to scrambler key. This has a two-folded meanings: a slight key change should lead to significant change in scrambled images during scrambling process, and also significant change in descrambled images during descrambling process. Fig. \ref{fig:sensitivity} shows the key sensitivity of \textit{Sudoku Associated Image Scrambler}, with two keys $K_a$ and $K_b$ differentiate from each other only for one bit. As can be seen, one bit change in scrambling key leads to two different scrambled images, whose difference image is also random-like. And descrambling using an incorrect key which is just one bit different from the correct key leads to random-like image.
\begin{eqnarray*}
  K_a &=& \textrm{B697F2703EA4347A85D997FB18A1FC3CE7E6901B6A9AE5EA} \\
  K_b &=& \textrm{A697F2703EA4347A85D997FB18A1FC3CE7E6901B6A9AE5EA}
\end{eqnarray*}

\section{Conclusion}
In this article, we mainly discussed the Sudoku associated two dimensional bijections and applications for image scrambling. We showed that these bijections are can be defined by Sudoku associated matrix element representations, which provide additional and parametric means to denote matrix elements besides the conventional way of using the row-column pair. Specifically, these new Sudoku associated matrix element representations are row-digit pair, digit-row pair, column-digit pair, digit-column pair, block-digit pair, and digit-block pair.

Since all these Sudoku associated matrix element representations are parametric with respect to a reference Sudoku matrix, it then allows us to denote matrix elements in secret ways and further provides Sudoku matrix dependent two dimensional bijections constructed by mapping from one representation to the other. We showed that many of these Sudoku associated two dimensional bijections have deterministic scrambling effect when we use them for image scrambling. For example, the bijection mapping from row-column pair to row-digit pair is equivalent to scramble pixels within a row to different positions that none two pixels that originally lies in the same column is still in the same column after scrambling; the bijection mapping from block-grid pair to block-digit pair is equivalent to scramble pixels within each block and cause a mosaic-like effect.

Furthermore, we proposed \textit{Sudoku Associated Image Scrambler}, a simple but effect digital image scrambler of using these Sudoku associated two dimensional bijections, by using a scrambling key to  control these bijections in a parametric way. Because a multiround scrambler is mathematically equivalent to a new bijection of composing a series of bijections in each scrambler round, we showed that these fundamental Sudoku associated two dimensional bijections can be cascaded together to scramble image pixels in a deterministic way. Simulation results of the proposed image scrambler and comparisons to recent peer algorithms indicate that \textit{Sudoku Associated Image Scrambler} outperforms or at least reaches state-of-the-art suggested by these peer algorithms \cite{ye2007image,ye2008novel,Abu,ye2010image} with respect to a number of evaluation and analysis methods, including human visual inspections, gray degree of scrambling, and autocorrelation coefficient of adjacent pixels. Moreover, statistical tests also support that \textit{Sudoku Associated Image Scrambler} does break the strong correlations between adjacent pixels to zero correlations after scrambling.

Similar scrambling ideas are definitely applicable to other digital formats such as digital audio and video. Since data hiding, data encryption, digital watermarking also widely use parametric bijections, we believe the proposed Sudoku associated two-dimensional bijections could also be useful in these areas.

\ifCLASSOPTIONcaptionsoff
  \newpage
\fi

\bibliographystyle{IEEEtran}
\bibliography{SudokuScrambler}
\end{document}